\documentclass[nofootinbib,notitlepage,onecolumn,secnumarabic,amssymb, nobibnotes, aps, prd]{revtex4-1}
\pdfoutput=1
\usepackage{graphicx,epsfig,youngtab} 
\usepackage{upgreek}
\usepackage{bbold}
\usepackage{amsmath}
\usepackage{ amssymb }
\usepackage{cancel}
\usepackage{amsfonts}
\usepackage{xcolor}
\usepackage{hyperref}
\usepackage{bm}
\usepackage{epstopdf}
\usepackage{natbib}
\usepackage{hyperref}

\newcommand{\cH}{\mathcal{H}}
\newcommand{\n}{{\bm {n}}}
\newcommand{\ud}{{\mathrm{d}}}
\newcommand{\p}{\partial}
\newcommand{\Q}{\mathcal{Q}}

\newcommand{\two}{^{{ ({{2}})}}}

\def\n{\bm{n}}
\def\x{\bm{x}}
\def\k{\bm{k}}
\def\be{\begin{equation}}
\def\ee{\end{equation}}
\def\bea{\begin{eqnarray}}
\def\eea{\end{eqnarray}}

\newcommand{\red}[1]{{\color{black}{#1}}}

\begin{document}

\title{\large Imprints of local lightcone projection effects on the galaxy bispectrum. {III}\\
\red{Relativistic corrections from nonlinear dynamical evolution on large-scales}}
\author{\large  Sheean Jolicoeur$^a$, Obinna Umeh$^{a,b}$,  Roy Maartens$^{a,b}$ and Chris Clarkson$^{c,a,d}$ 
\\~\\
\emph{\normalsize $^a$Department of Physics \& Astronomy, University of the Western Cape,
Cape Town 7535, South Africa \\
$^b$Institute of Cosmology \& Gravitation, University of Portsmouth, Portsmouth PO1 3FX, United Kingdom\\
$^c$School of Physics \& Astronomy, Queen Mary University of London, London E1 4NS, United Kingdom\\
$^d$Department of Mathematics \& Applied Mathematics, University of Cape Town, Cape Town 7701, South Africa}~\\~\\}

\date{\today}

\begin{abstract}

 The galaxy bispectrum is affected on equality scales and above by relativistic observational effects, at linear and nonlinear order. These lightcone effects include  local contributions from Doppler and gravitational potential terms, as well as   integrated contributions like lensing, together with all the couplings at nonlinear order. We recently presented the correction to the  galaxy bispectrum  from all local lightcone effects up to second order in perturbations, \red{using a plane-parallel approximation}. Here we update our previous result by including the effects from relativistic nonlinear dynamical evolution. \red{We show that these dynamical effects make a significant contribution to the projection effects.}
\par

\end{abstract}

\maketitle

\section{Introduction}

Galaxy counts are distorted due to the fact that we observe on the past lightcone. The Kaiser redshift-space distortion (RSD) effect is well known. Lensing convergence also distorts the number counts. There are further distortions from Doppler, Sachs-Wolfe, ISW and time-delay effects, which are suppressed on sub-equality scales. At nonlinear order, there are couplings amongst all of these projection effects.

The full set of relativistic projection effects has been computed at first order by \cite{Yoo:2010ni,Challinor:2011bk,Bonvin:2011bg}, and then at second order by \cite{Bertacca:2014dra,Bertacca:2014wga,Yoo:2014sfa,DiDio:2014lka,Bertacca:2014hwa}. The observed galaxy power spectrum at tree level involves only the first-order projection effects. On super-equality scales, the relativistic projection effects are similar to the effects of scale-dependent bias, so that these effects need to be taken into account when measuring or constraining primordial non-Gaussianity~\cite{Bruni:2011ta,Jeong:2011as,Camera:2014bwa,Camera:2014sba}.

The observed galaxy bispectrum at tree level involves both first- and second-order projection effects, which on ultra-large scales could also be mistaken for primordial non-Gaussianity. We recently computed the galaxy bispectrum with all local relativistic projection effects, \red{i.e., neglecting terms involving lensing and other line-of-sight integrals.
This approximation is imposed on us since,  
in the early papers of this series, we work in Fourier space (in common with much of the literature on the galaxy bispectrum). 
In principle it is known how to include the integrated contributions -- by working with the relativistic galaxy three-point correlation function or angular harmonic bispectrum. But in practice this is a formidable challenge which has not yet been met.
Another approximation that is imposed by a standard Fourier-space analysis is the plane-parallel approximation, which excludes wide-angle correlations. Wide-angle correlations would be automatically included in an angular harmonic or three-point correlation analysis that inorporated integrated effects.
We plan to include integrated and wide-angle contributions in future papers of this series. } 

Papers I and II provided a compact kernel for easy Fourier-space computations: \red{in Paper I~\cite{Umeh:2016nuh} we presented the main results and in Paper II~\cite{Jolicoeur:2017nyt} we gave the details behind the main results, and also generalised these results.} (See also other computations for special cases in~\cite{Kehagias:2015tda,DiDio:2015bua,DiDio:2016gpd} and a formalism for the general case in~\cite{Bertacca:2017dzm}.) 

\red{In this Paper III of the series, we include additional effects (identified in Paper II) to update the general result in Paper II. We do not repeat the general treatment of Paper II, but refer the reader to Paper II for the general discussion, the detailed derivations of key results and their full expressions. For convenience, we include in Appendix~\ref{AA} some of the relevant basic formulas and definitions.}
 
\red{The expression for the observed galaxy number count contrast at second order \cite{Bertacca:2014dra,Bertacca:2014wga,Yoo:2014sfa,DiDio:2014lka,Bertacca:2014hwa} is extremely long and complicated -- even if we omit all terms with integrated contributions. A convenient form in the local case is given in Paper II~\cite{Jolicoeur:2017nyt}, in the Poisson gauge. In Paper III our focus is on the contributions from the second-order gravitational potentials and the peculiar velocity potential: we used the standard (Newtonian) expressions in Paper II and here we will compute the relativistic corrections. These contributions enter the observed galaxy number contrast as follows (see Paper II~\cite{Jolicoeur:2017nyt} for the full expression)}:
\bea
\Delta_g^{(2)}&=& \delta_g^{(2)}\nonumber\\
&&{}- \frac{1}{\mathcal{H}}\partial_{\parallel}^{2}v^{(2)}+ \bigg[b_{e} - 2\mathcal{Q} -\frac{2(1-\mathcal{Q})}{\chi\mathcal{H}}- \frac{\mathcal{H}'}{\mathcal{H}^{2}} \bigg]\left[\partial_{\parallel}v^{(2)}-\Phi^{(2)} \right] +2(\mathcal{Q}-1)\Psi^{(2)}  +\Phi^{(2)} + \frac{1}{\mathcal{H}}{\Psi^{(2)\prime}} \nonumber \\
\label{odg2}
&&{}+\mbox{very many terms quadratic in first-order  quantities,}
\eea
\red{where $\Delta_g^{(2)}(z,\n)$ is the second-order number count contrast at observed redshift $z$ and in the observed direction $\n$.
The gravitational potentials $\Phi,\Psi$ in the perturbed metric and the peculiar velocity potential $v$ are defined in Appendix \ref{AA}. Here $\partial_{\parallel}=n^i\partial_i$ is the line of sight derivative, the conformal Hubble rate is $\cH=(\ln a)'$, where a prime denotes the conformal time derivative, and $\chi$ is the line-of-sight comoving distance. The astrophysical quantities $b_e$ and ${\cal Q}$ are the evolution bias and magnification bias of the galaxy distribution, defined in Appendix~\ref{AA}.}

\red{We first consider the intrinsic second-order number contrast $\delta_g^{(2)}$, which is related to the matter density contrast via a galaxy bias model. In Paper II~\cite{Jolicoeur:2017nyt} we presented a careful treatment of galaxy bias that is relativistic, i.e., gauge-independent and hence consistent on ultra-large scales. In order to achieve this, we used the simplest possible model of galaxy bias on sub-Hubble scales, i.e. the Eulerian local-in-mass-density model~\cite{Desjacques:2016bnm}. Even for this simple model, the gauge-independent extension to general relativity is highly non-trivial. It leads to a  complicated expression for the Poisson-gauge number contrast in terms of the density contrast $\delta_{\rm T}$ in the total-matter gauge~\cite{Jolicoeur:2017nyt}. (Note that T gauge corresponds to an Eulerian frame.) The expression that we derived in Paper II is of the form:
\bea
\delta_{g }^{{{({2})}}}&=&b_1 \delta^{(2)}_{\rm T}+ b_2\big[\delta^{(1)}_{\rm T}\big]^2+(3-b_{e})\mathcal{H}v^{(2)}+\mbox{many terms quadratic in first-order  quantities,} \label{dg2b}
\eea
where $b_1, b_2$ are the Eulerian bias coefficients.}
In \eqref{dg2b}, the term $(3-b_{e})\mathcal{H}v^{(2)}$ and the quadratic terms keep the bias relation gauge-independent, i.e. consistent with GR. (See Paper II~\cite{Jolicoeur:2017nyt} for details, and for the
full expressions.)

In Papers I and II~\cite{Umeh:2016nuh,Jolicoeur:2017nyt}, \red{we focused on the relativistic local projection effects, and for simplicity we used the standard Newtonian approximations to compute the nonlinear dynamical evolution of the potentials}~\cite{Bernardeau:2001qr}:
\begin{eqnarray}
\red{\delta^{(2)}_{\rm T}}(\bm{k}_3) &\approx& \red{\delta^{(2)}_{\rm T\,N}}(\bm{k}_3) = \int \ud(\bm{k}_{1}, \bm{k}_{2}, \bm{k}_{3}) \,  {F_2}(\bm{k}_{1},\bm{k}_{2}),  
\label{nd2} \\
 v^{{{({2})}}}(\bm{k}_3) &\approx& \red{v^{{{({2})}}}_{\rm N}}(\bm{k}_3) = f\,\frac{\cH}{k^2_3} \int \ud(\bm{k}_{1}, \bm{k}_{2}, \bm{k}_{3})\,G_2(\bm{k}_{1},\bm{k}_{2})
, \label{nv2}
\\
\Phi^{{{({2})}}}(\k_3) &\approx& \red{\Phi^{{{({2})}}}_{\rm N}}(\k_3) =-{3\over2}\Omega_m\frac{\mathcal{H}^2 }{k^2_3} \,\red{\delta^{{{({2})}}}_{\rm T\,N}}(\bm{k}_3),
\label{np2}\\
    \Psi^{{{({2})}}}(\bm{k}_3)&\approx& \red{\Psi^{{{({2})}}}_{\rm N}}(\k_3)=\Phi^{{{({2})}}}_{\rm N}(\k_3),
\label{np3}
\end{eqnarray}
where  \red{$f$ is the growth rate of matter perturbations, defined in \eqref{e5a} below, the kernels $F_2$, $G_2$ are derived in Appendix~\ref{AA},} and
\begin{equation}
\int \ud(\bm{k}_{1}, \bm{k}_{2}, \bm{k}_{3}) \equiv \int\frac{\ud \k_{1}}{(2\pi)^{3}}\int\ud \k_{2}\,\red{\delta^{(1)}_{\rm T}}(\bm{k}_{1})\,\red{\delta^{(1)}_{\rm T}}(\bm{k}_{2})\,\delta^{D}(\bm{k}_{1}+\bm{k}_{2}-\bm{k}_{3}). \label{e14}
\end{equation}

\red{To begin, we deal with a critical subtlety involving 
the dark matter density contrast $\delta^{(2)}_{\rm T}$, which has a relativistic correction $\delta^{(2)}_{\rm T\,GR}$ of order $(\cH^2/k^2) [\delta^{(1)}_{\rm T}]^2$ from the field equations (see~\cite{Villa:2015ppa}).
In the bias relation \eqref{dg2b},
it is implicit that $\delta^{(2)}_{\rm T}$ is smoothed on a fixed  physical scale $R$ corresponding to fixed-mass halo formation: 
\be \label{dg2b2}
\delta_{g }^{(2)}=b_1 \,\delta^{(2)}_{\rm T}\Big|_{R}+ b_2\left[\,\delta^{(1)}_{\rm T}\Big|_{R}\right]^2+ \mbox{terms to enforce gauge-independence.} 
\ee
If the primordial metric perturbation is Gaussian, then the small-scale density at a fixed local physical scale is independent of the long-wavelength mode $\delta^{(2)}_{\rm T\,GR}$: 
\be \label{dg2b3}
\delta^{(2)}_{\rm T}\Big|_{R}= \delta^{(2)}_{\rm T\, N}\Big|_{R}.
\ee
For the local observer at the galaxy, the long mode has no effect.
(See \cite{Dai:2015jaa, dePutter:2015vga, Bartolo:2015qva} for detailed arguments.)} 
Therefore \red{$\delta^{(2)}_{\rm T\, GR}$} only affects the dark matter bispectrum, but {\em not} the galaxy bispectrum, and we do not need it for the relativistic dynamical corrections. \red{We emphasize that this is a relativistic requirement -- the long mode $\delta^{(2)}_{\rm T\, GR}$ cannot enter the bias relation if we are to retain coordinate invariance in the local patch.}

However we do need the relativistic corrections to the velocity and gravitational potentials.
These horizon-scale corrections from the field equations are of order $(\cH^4/k^4) [\delta^{(1)}_{\rm T}]^2$, as shown in \eqref{e2}--\eqref{e4} below.

In the following sections, we calculate the relativistic dynamical corrections to the second-order velocity and gravitational potentials, 
present the updated kernel for the galaxy bispectrum, and compute the changes to the galaxy bispectrum introduced by the dynamical corrections.

\section{Relativistic dynamical corrections to the bispectrum kernel}

\red{In Fourier space, the observed galaxy bispectrum $B_g$ at fixed redshift is given by
\begin{equation} 
\big\langle \Delta_{g}( z,\bm{k}_{1}) \Delta_{g}( z,\bm{k}_{2}) \Delta_{g}( z,\bm{k}_{3})\big \rangle = (2\pi)^{3}B_{g}( z, \bm{k}_{1},  \bm{k}_{2},  \bm{k}_{3})\delta^{D}( \bm{k}_{1}+ \bm{k}_{2}+ \bm{k}_{3}).
\end{equation}
From now on we will suppress the redshift dependence for brevity.  At second order, the only combinations of terms that contribute at tree level  are
\be 
\big \langle \Delta_{g}( \bm{k}_{1}) \Delta_{g}( \bm{k}_{2}) \Delta_{g}( \bm{k}_{3})\big \rangle = {1\over2}
\big\langle \Delta_{g}^{{{({1})}}}(\bm{k}_{1}) \Delta_{g}^{{{({1})}}}(\bm{k}_{2}) \Delta_{g}^{{{({2})}}}( \bm{k}_{3})\big \rangle  + \text{2\;cyc.\;perm.,}
\ee
and using Wick's theorem, this leads to}~\cite{Jolicoeur:2017nyt}
\be
B_{g}( \bm{k}_{1},  \bm{k}_{2},  \bm{k}_{3}) = \mathcal{K}^{{{({1})}}}(\bm{k}_{1})\, \mathcal{K}^{{{({1})}}}(\bm{k}_{2}) \,\mathcal{K}^{{{({2})}}}(\bm{k}_{1},  \bm{k}_{2},\bm{k}_{3})\, P(k_{1})P(k_{2}) +  \text{2 cyc. perm.},
\ee
\red{where $P$ is the power spectrum of $\delta^{(1)}_{\rm T}$.} 

The first-order kernel has Newtonian and relativistic parts~\red{\cite{Jeong:2011as}}:
\be
\mathcal{K}^{{{({1})}}}=\mathcal{K}^{{{({1})}}}_{\rm{N}}+\mathcal{K}^{{{({1})}}}_{\rm{GR}},\qquad
\mathcal{K}^{{{({1})}}}_{\rm{N}}({\k})=   b_{1} + f\mu^{2}\,,\qquad  {
\mathcal{K}^{{{({1})}}}_{\rm{GR}}({\k}) ={\rm i} \,\mu\, {\gamma_1\over k}+\frac{\gamma_2}{k^{{2}}} , } \qquad \mu =  \hat{\bm{k}} \cdot {\bm{n}}\,, \label{k1}
\ee
\red{where the coefficients $\gamma_a(z)$ are given in Appendix~\ref{AA}. Since the observed direction $\n$ is fixed, we are necessarily assuming a plane-parallel approximation.}

The standard Newtonian part of the second-order kernel is~\red{\cite{Verde:1998zr,Scoccimarro:1999ed}}
\begin{eqnarray}\label{eq:FourierNewtonian}
{\mathcal{K}^{{{({2})}}}_{\rm{N} }(\bm{k}_{1}, \bm{k}_{2},{\k_3})} &=& b_{1}F_{2}(\bm{k}_{1}, \bm{k}_{2}) + b_{2} + fG_{2}(\bm{k}_{1}, \bm{k}_{2})\mu_{3}^{2}
\nonumber\\ &&{}
+  f^2{\mu_1\mu_2 \over k_1k_2}\big( \mu_1k_1+\mu_2k_2\big)^2
+
b_1{f\over k_1k_2}\Big[ \big(\mu_1^2+\mu_2^2 \big)k_1k_2+\mu_1\mu_2\big(k_1^2+k_2^2 \big) \Big],
\end{eqnarray}
where \red{$\mu_a=\hat{\bm{k}}_a \cdot {\bm{n}}$}. The \red{$G_2$} term is the second-order Kaiser RSD term, and the terms in the second line are other nonlinear RSD contributions.

The relativistic part of the kernel is derived in Paper II~\cite{Jolicoeur:2017nyt}:
\begin{eqnarray}\label{k2g}
\mathcal{K}^{{{({2})}}}_{\mathrm{GR}}(\bm{k}_{1}, \bm{k}_{2}, \bm{k}_{3}) &=& \frac{1}{k_{1}^{2}k_{2}^{2}}\bigg\{\Gamma_{1} + {\rm i}\left(\mu_{1}k_{1} + \mu_{2}k_{2}\right)\Gamma_{2} + \frac{k_{1}^2k_{2}^2}{k_{3}^2} \Big[F_{2}(\bm{k}_{1}, \bm{k}_{2})\,\Gamma_{3} + G_{2}(\bm{k}_{1}, \bm{k}_{2})\,\Gamma_{4} \Big] \nonumber \\&&{}
 + \left(\mu_{1}\mu_{2}k_{1}k_{2}\right)\Gamma_{5} + \left(\bm{k}_{1}\cdot \bm{k}_{2}\right)\Gamma_{6} + \left(k_{1}^{2} + k_{2}^{2}\right)\Gamma_{7} + \left(\mu_{1}^{2}k_{1}^{2} + \mu_{2}^{2}k_{2}^{2}\right)\Gamma_{8} 
 \nonumber \\&&{}
 +  {\rm i}\bigg[\left(\mu_{1}k_{1}^{3} + \mu_{2}k_{2}^{3}\right)\Gamma_{9}  
+ \left(\mu_{1}k_{1} + \mu_{2}k_{2}\right)\left(\bm{k}_{1} \cdot \bm{k}_{2}\right)\Gamma_{10}  + k_{1}k_{2}\left(\mu_{1}k_{2} + \mu_{2}k_{1}\right)\Gamma_{11}   
\nonumber \\&&{}
~~~~+ \left(\mu_{1}^{3}k_{1}^{3}+ \mu_{2}^{3}k_{2}^{3}\right)\Gamma_{12}
 + \mu_{1}\mu_{2}k_{1}k_{2}\left(\mu_{1}k_{1} + \mu_{2}k_{2}\right)\Gamma_{13} + \mu_{3}\frac{k_{1}^{2}k_{2}^{2}}{k_{3}}\,G_{2}(\bm{k}_{1}, \bm{k}_{2})\,\Gamma_{14}\bigg] \bigg\}\,,
\end{eqnarray}
where
the $\Gamma_I(z)$ are given in~\cite{Jolicoeur:2017nyt}. They are ordered  according to the powers of $\cH/k$, starting with the ${\cal O}(\cH^4/k^4)$ term and ending with the ${\cal O}(\cH/k)$ terms.

Equation \eqref{k2g} was derived in Paper II using the standard expressions \eqref{nv2}--\eqref{np3} for the potentials.
These Newtonian forms need to be corrected by the full relativistic expressions, \red{$v\two=v_{ \mathrm{N} }\two+v_{ \mathrm{GR} }\two$ and similarly for $\Phi\two$ and $\Psi\two$.} For $\Lambda$CDM and Gaussian initial conditions, the relativistic parts, in real space and Poisson gauge,  are~\cite{Villa:2015ppa}:  
\begin{eqnarray}
v_{ \mathrm{GR} }\two(\bm{x})  &=&  \alpha D'g \left[ \Big(1 - \frac{10}{3} {g_{\rm in}\over g} \Big) \varphi_0^2(\bm{x})  - 12\, \Theta_0 (\bm{x})\right], \label{e2} \\
\Phi_{ \mathrm{GR} }\two(\bm{x}) &=& \left( 3 g^2 - \frac 5 3 g g_{\rm in}   + \frac{ \alpha D^{\prime 2}}{a}  \right)\varphi_0^2(\bm{x}) 
+ 12 \left(2 g^2 - \frac{5}{3} g g_{\rm in} + \frac{ \alpha D^{\prime 2}}{a} \right) \Theta_0(\bm{x}), \label{e3} \\
\Psi_{ \mathrm{GR} }\two(\bm{x}) &=& - \left( g^2 + \frac 5 3 g g_{\rm in} - \frac{ \alpha D^{\prime 2}}{a}  \right)\varphi_0^2(\bm{x}) 
+ 12\left(  g^2 -\frac{5}{3}g g_{\rm in}  \right) \Theta_{0}(\bm{x}). \label{e4}
\end{eqnarray}
Here 
\bea
\alpha&=& {2\over 3\Omega_{m0}H_0^2}= {2\over 3\Omega_{m}\cH^2a}.\label{e0}
\\
g\varphi_0&=& \Psi^{(1)}=\Phi^{(1)},   \quad g={D\over a},\quad D(\eta)={\red{\delta^{(1)}_{\rm T}}(\eta,\x)\over \red{\delta^{(1)}_{\rm T}}(\eta_0,\x)} , \label{e1}\\
 \Theta_{0}(\bm{x}) &=& \frac{1}{2} \nabla^{-2}  \bigg[ \frac 1 3 \varphi_0^{,i}  \varphi_{0,i} -  \nabla^{-2} \Big( \varphi_0^{,i}  \varphi_{0}^{,j} \Big)_{,ij} \bigg], \label{e5}\\
{g_{\rm in}\over g}&=&{1\over5}\Big(3+2{f\over\Omega_m}\Big),\quad \red{f={\ud \ln D \over \ud \ln a}}, 
\label{e5a}
\eea
\red{where $g_{\rm in}$ is the initial value in the matter-dominated era, 0 denotes redshift $z=0$ and $a_0=D_0=g_0=1$.} Now we simplify \eqref{e2}--\eqref{e4} using \eqref{e0}--\eqref{e5a}:
\begin{eqnarray}
\cH v_{ \mathrm{GR} }\two(\bm{x}) &=&- g^2\,\frac{2f}{3\Omega_m}\bigg[\Big(1 +\frac{4f}{3\Omega_{m} }\Big)\varphi_0^2(\bm{x})+ 12\, \Theta_0 (\bm{x})\bigg],\label{e6}\\
\Phi_{ \mathrm{GR} }\two(\bm{x}) &=& g^{2}\bigg[2 -\frac{2f}{3\Omega_m}+ \frac{2f^{2}}{3\Omega_{m}}\bigg]\varphi_0^2(\bm{x}) + 12g^{2}\bigg[1-\frac{2f}{3\Omega_m}+\frac{2f^{2}}{3\Omega_{m}}\bigg]\Theta_0 (\bm{x}), \label{e7} \\ 
\Psi_{ \mathrm{GR} }\two(\bm{x})&=& -g^{2}\bigg[2 + \frac{2f}{3 \Omega_m}  - \frac{2f^{2}}{3\Omega_{m}}\bigg]\varphi_0^2(\bm{x}) - 8g^{2}\,\frac{f}{\Omega_m}\,\Theta_0 (\bm{x}). \label{e8}
\end{eqnarray}

The matter density contrast  is given by 
\be
 \red{\delta^{(1)}_{\rm T}}=\alpha D \nabla^2 \varphi_0.
\ee
Together with \eqref{e1} and \eqref{e5}, this leads to the Fourier transforms: 
\begin{eqnarray}
\varphi_0^2(\bm{k}_{3}) &=& \bigg(\frac{3\Omega_{m}\cH^{2}}{2g}\bigg)^{2}\int \ud(\bm{k}_{1}, \bm{k}_{2}, \bm{k}_{3})\, \frac{1}{k_{1}^{2}k_{2}^{2}}, \label{e10} \\
\Theta_0(\bm{k}_{3}) &=&  \bigg(\frac{3\Omega_{m}\cH^{2}}{2gk_{3}}\bigg)^{2}\int \ud(\bm{k}_{1}, \bm{k}_{2}, \bm{k}_{3})\,\bigg\{\frac{\bm{k}_{1}\cdot \bm{k}_{2}}{6k_{1}^{2}k_{2}^{2}} - \frac{1}{2k_{3}^{2}}\bigg[1+\frac{\bm{k}_{1}\cdot \bm{k}_{2}}{k_{1}k_{2}}\bigg(\frac{k_{1}}{k_{2}} + \frac{k_{2}}{k_{1}}\bigg) + \frac{(\bm{k}_{1}\cdot \bm{k}_{2})^{2}}{k_{1}^{2}k_{2}^{2}}\bigg]\bigg\}. \label{e13}
\end{eqnarray}
Then \eqref{e6}--\eqref{e8} become 
\begin{eqnarray}
\cH v_{ \mathrm{GR} }\two(\bm{k}_{3}) &=& 3\Omega_{m}\cH^{4} f \int \ud(\bm{k}_{1}, \bm{k}_{2}, \bm{k}_{3})\,\frac{1}{k_{1}^{2}k_{2}^{2}}\bigg[-\frac{1}{6}\bigg(3+\frac{4f}{\Omega_{m}}\bigg)  + 
E_{2}(\bm{k}_{1}, \bm{k}_{2},\k_3)\bigg], \label{e15}\\
\Phi_{ \mathrm{GR} }\two(\bm{k}_{3}) &=& 3\Omega_{m}\cH^{4} \int \ud(\bm{k}_{1}, \bm{k}_{2}, \bm{k}_{3})\,\frac{1}{k_{1}^{2}k_{2}^{2}}\bigg[\frac{1}{2}\big(3\Omega_{m}-f{+f^{2}}\big) -{1\over2}\big(3\Omega_{m}-2f + 2f^2\big)            E_{2}(\bm{k}_{1}, \bm{k}_{2},\k_3)
\bigg],  \label{e17} \\
\Psi_{ \mathrm{GR} }\two(\bm{k}_{3}) &=& 3\Omega_{m}\cH^{4} \int \ud(\bm{k}_{1}, \bm{k}_{2}, \bm{k}_{3})\,\frac{1}{k_{1}^{2}k_{2}^{2}}\bigg[-\frac{1}{2}\big(3\Omega_{m}+f{-f^{2}}\big) +f\,
E_{2}(\bm{k}_{1}, \bm{k}_{2},\k_3)
\bigg]. \label{e18} 
\end{eqnarray}
Here we have defined a new kernel function, which scales as $k^0$, like $F_2$ and $G_2$:
\begin{equation}
E_{2}(\bm{k}_{1}, \bm{k}_{2},\k_3)=\frac{k_{1}^{2}k_{2}^{2}}{k_{3}^{4}}
\bigg[3+2\frac{\bm{k}_{1}\cdot \bm{k}_{2}}{k_{1}k_{2}}\bigg(\frac{k_{1}}{k_{2}} + \frac{k_{2}}{k_{1}}\bigg) + \frac{(\bm{k}_{1}\cdot \bm{k}_{2})^{2}}{k_{1}^{2}k_{2}^{2}}\bigg]. \label{e16}
\end{equation}
For the time derivative of \eqref{e18}, we find that
\bea
\Psi_{ \mathrm{GR} }^{(2) \prime}(\bm{k}_{3}) &=& 3\Omega_{m}\cH^{5}\int \ud(\bm{k}_{1}, \bm{k}_{2}, \bm{k}_{3})\,\frac{1}{k_{1}^{2}k_{2}^{2}}\bigg\{{1\over2}(1-f)\bigg[ 6\Omega_{m}+f(1-2f) -2f \frac{\cH'}{\cH^{2}}\bigg] 
+{1\over2}(2f-1){f'\over\cH}
\nonumber \\
&&{}\qquad \qquad \qquad \qquad \qquad\qquad\quad +\bigg[f\bigg(2f-1+\frac{2\cH'}{\cH^{2}}\bigg) + \frac{f'}{\cH}\bigg]
E_{2}(\bm{k}_{1}, \bm{k}_{2},\k_3)
\bigg\}.  
\label{e19}
\eea

From \eqref{e15}--\eqref{e19}, it follows that:
\begin{itemize}
\item
 The potential terms proportional to $v^{(2)}, \Phi^{(2)}, \Psi^{(2)}$ and $\Psi^{(2)\prime}$ in \eqref{odg2} and \eqref{dg2b}, lead to relativistic corrections proportional to  $v^{(2)}_{\rm GR}, \Phi^{(2)}_{\rm GR}, \Psi^{(2)}_{\rm GR}$ and $\Psi^{(2)\prime}_{\rm GR}$, which sum up to a term of the form\\ 

  \red{
 $ \big[\alpha_1(z)+E_{2}(\bm{k}_{1}, \bm{k}_{2},\k_3)\,\alpha_2(z)\big]/(k_1k_2)^2$.} 
 \item 
 For the Doppler term in \eqref{odg2}, proportional to $\partial_{\parallel}v^{(2)}$, the relativistic correction $\partial_{\parallel}v^{(2)}_{\rm GR}$ leads to a term of the form \\
 
\red{$\mu_3k_3\big[\alpha_3(z)+E_{2}(\bm{k}_{1}, \bm{k}_{2},\k_3)\,\alpha_4(z)\big]/(k_1k_2)^2$}. 
\item
For the second-order Kaiser term in \eqref{odg2}, $-{\mathcal{H}^{-1}}\partial_{\parallel}^{2}v^{(2)}$, the relativistic correction $-{\mathcal{H}^{-1}}\partial_{\parallel}^{2}v^{(2)}_{\rm GR}$ is of the form \\

\red{$\mu_3^2k_3^2\big[\alpha_5(z)+E_{2}(\bm{k}_{1}, \bm{k}_{2},\k_3)\,\alpha_6(z)\big]/(k_1k_2)^2$}. 
\end{itemize}

\red{Therefore the relativistic dynamical correction to the Paper II kernel \eqref{k2g} is determined by  6 new coefficients $\alpha_I$, and contains terms of order $(\cH/k)^n$, $n=2,3,4$.} The revised GR kernel with the relativistic dynamical corrections is given by
\begin{eqnarray}\label{e26}
\mathcal{K}^{(2)}_{\mathrm{GR}}(\bm{k}_{1}, \bm{k}_{2}, \bm{k}_{3}) &=& \frac{1}{k_{1}^{2}k_{2}^{2}}\Bigg\{\Gamma_{1} + {\alpha_1+E_{2}(\bm{k}_{1}, \bm{k}_{2},\k_3)\,\alpha_2} + {\rm i}\bigg[\left(\mu_{1}k_{1} + \mu_{2}k_{2}\right)\Gamma_{2}  + {\mu_{3}k_{3}{\Big(\alpha_3+E_{2}(\bm{k}_{1}, \bm{k}_{2},\k_3)\,\alpha_4 \Big)}}\bigg] \nonumber \\
&&{} \qquad   + \frac{k_{1}^2k_{2}^2}{k_{3}^2} \Big[F_{2}(\bm{k}_{1}, \bm{k}_{2})\,\Gamma_{3} + G_{2}(\bm{k}_{1}, \bm{k}_{2})\,\Gamma_{4} \Big] + \left(\mu_{1}k_{1}\mu_{2}k_{2}\right)\Gamma_{5} +
{\mu_{3}^{2}k_{3}^{2}{\Big(\alpha_5+E_{2}(\bm{k}_{1}, \bm{k}_{2},\k_3)\,\alpha_6 \Big)}} \nonumber \\
&&{} \qquad + \left(\bm{k}_{1}\cdot \bm{k}_{2}\right)\Gamma_{6} + \left(k_{1}^{2} + k_{2}^{2}\right)\Gamma_{7} + \left(\mu_{1}^{2}k_{1}^{2} + \mu_{2}^{2}k_{2}^{2}\right)\Gamma_{8} + {\rm i}\bigg[\left(\mu_{1}k_{1}^{3} + \mu_{2}k_{2}^{3}\right)\Gamma_{9}  
 \nonumber \\
 &&{} \qquad + \left(\mu_{1}k_{1} + \mu_{2}k_{2}\right)\left(\bm{k}_{1} \cdot \bm{k}_{2}\right)\Gamma_{10} + k_{1}k_{2}\left(\mu_{1}k_{2} + \mu_{2}k_{1}\right)\Gamma_{11} + \left(\mu_{1}^{3}k_{1}^{3}+ \mu_{2}^{3}k_{2}^{3}\right)\Gamma_{12}
  \nonumber \\
 &&{} \qquad + \mu_{1}\mu_{2}k_{1}k_{2}\left(\mu_{1}k_{1} + \mu_{2}k_{2}\right)\Gamma_{13} + \mu_{3}\frac{k_{1}^{2}k_{2}^{2}}{k_{3}}\,G_{2}(\bm{k}_{1}, \bm{k}_{2})\,\Gamma_{14}\bigg] \Bigg\}.
\end{eqnarray}
The new redshift-dependent functions $\alpha_{I}$ are given in Appendix \ref{A}. 

Equation \eqref{e26} shows that $\alpha_1$ is actually a correction to $\Gamma_1$, whereas the remaining $\alpha_I$ are new coefficients. This means that the revised GR kernel has a total of $14+6-1=19$ independent coefficients, which we re-label as follows:
\be\label{beta}
\Gamma_1+\alpha_1,\, \alpha_2,\, \cdots,\, \Gamma_{14} \quad \to \quad {\beta_1,\, \cdots, \,\beta_{19}}\,.
\ee
\red{(The $\beta_I$ should not be confused with the RSD parameter $\beta=f/b_1$.)}
Using this re-labelling, the revised kernel \eqref{e26} is rewritten as:
\begin{eqnarray}\label{e26r}
\mathcal{K}^{(2)}_{\mathrm{GR}}(\bm{k}_{1}, \bm{k}_{2}, \bm{k}_{3}) &=& \frac{1}{k_{1}^{2}k_{2}^{2}}\Bigg\{\beta_{1} + E_{2}(\bm{k}_{1}, \bm{k}_{2},\k_3)\,\beta_2 + {\rm i}\bigg[\left(\mu_{1}k_{1} + \mu_{2}k_{2}\right)\beta_{3}  + {\mu_{3}k_{3}{\Big(\beta_4+E_{2}(\bm{k}_{1}, \bm{k}_{2},\k_3)\,\beta_5 \Big)}}\bigg] \nonumber \\
&&{} \qquad   + \frac{k_{1}^2k_{2}^2}{k_{3}^2} \Big[F_{2}(\bm{k}_{1}, \bm{k}_{2})\,\beta_{6} + G_{2}(\bm{k}_{1}, \bm{k}_{2})\,\beta_{7} \Big] + \left(\mu_{1}k_{1}\mu_{2}k_{2}\right)\beta_{8} +
{\mu_{3}^{2}k_{3}^{2}{\Big(\beta_9+E_{2}(\bm{k}_{1}, \bm{k}_{2},\k_3)\,\beta_{10} \Big)}} \nonumber \\
&&{} \qquad + \left(\bm{k}_{1}\cdot \bm{k}_{2}\right)\beta_{11} + \left(k_{1}^{2} + k_{2}^{2}\right)\beta_{12} + \left(\mu_{1}^{2}k_{1}^{2} + \mu_{2}^{2}k_{2}^{2}\right)\beta_{13} 
\nonumber \\
 &&{} \qquad+ {\rm i}\bigg[\left(\mu_{1}k_{1}^{3} + \mu_{2}k_{2}^{3}\right)\beta_{14}  
  + \left(\mu_{1}k_{1} + \mu_{2}k_{2}\right)\left(\bm{k}_{1} \cdot \bm{k}_{2}\right)\beta_{15} + k_{1}k_{2}\left(\mu_{1}k_{2} + \mu_{2}k_{1}\right)\beta_{16} 
  \nonumber \\
 &&{} \qquad\qquad + \left(\mu_{1}^{3}k_{1}^{3}+ \mu_{2}^{3}k_{2}^{3}\right)\beta_{17}+ \mu_{1}\mu_{2}k_{1}k_{2}\left(\mu_{1}k_{1} + \mu_{2}k_{2}\right)\beta_{18} + \mu_{3}\frac{k_{1}^{2}k_{2}^{2}}{k_{3}}\,G_{2}(\bm{k}_{1}, \bm{k}_{2})\,\beta_{19}\bigg] \Bigg\}.
\end{eqnarray}
For convenience, we give in Appendix~\ref{B} the explicit forms for all the $\beta_I$, \red{which replace the $\Gamma_I$ of Paper II.}

\section{Computing the galaxy bispectrum}

\begin{figure}[! ht]
\centering
\includegraphics[width=0.32\textwidth] {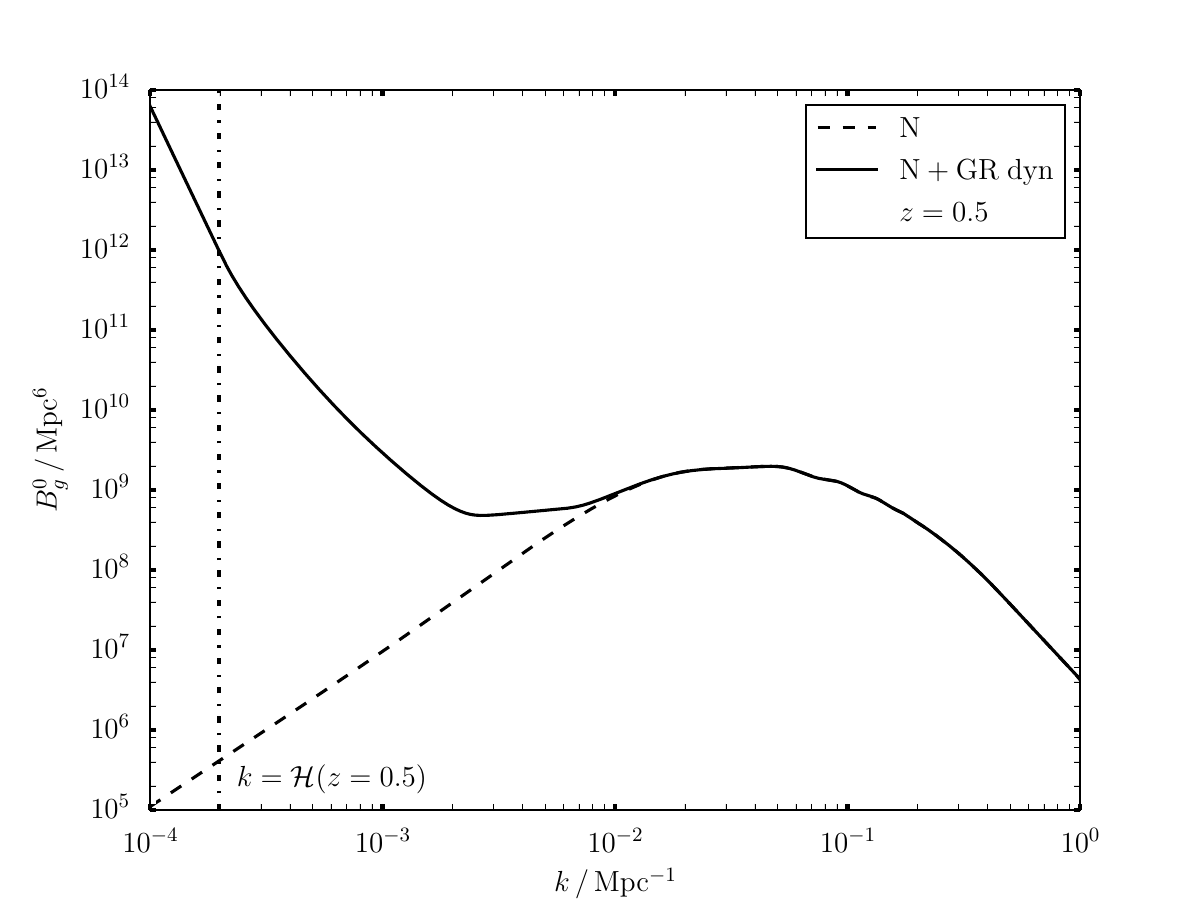}
\includegraphics[width=0.32\textwidth] {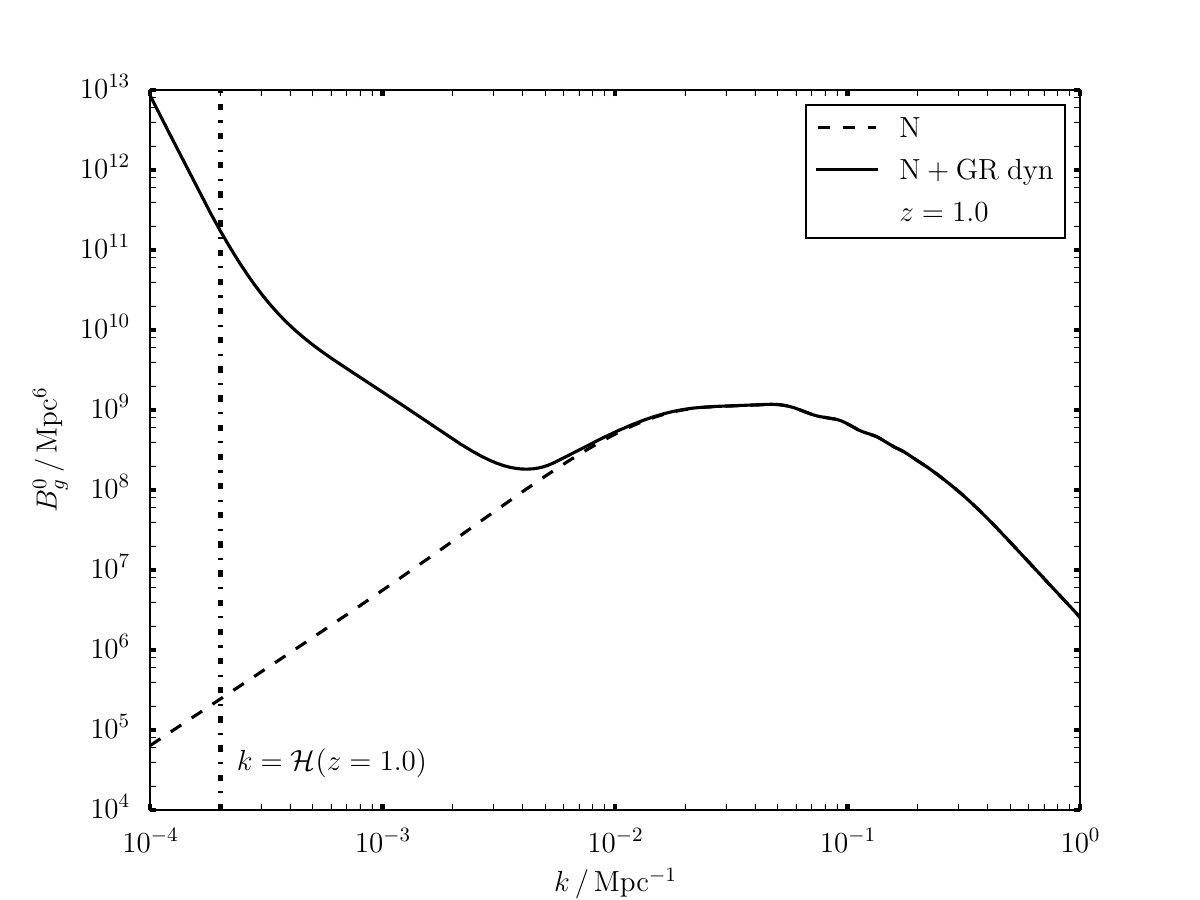}
\includegraphics[width=0.32\textwidth] {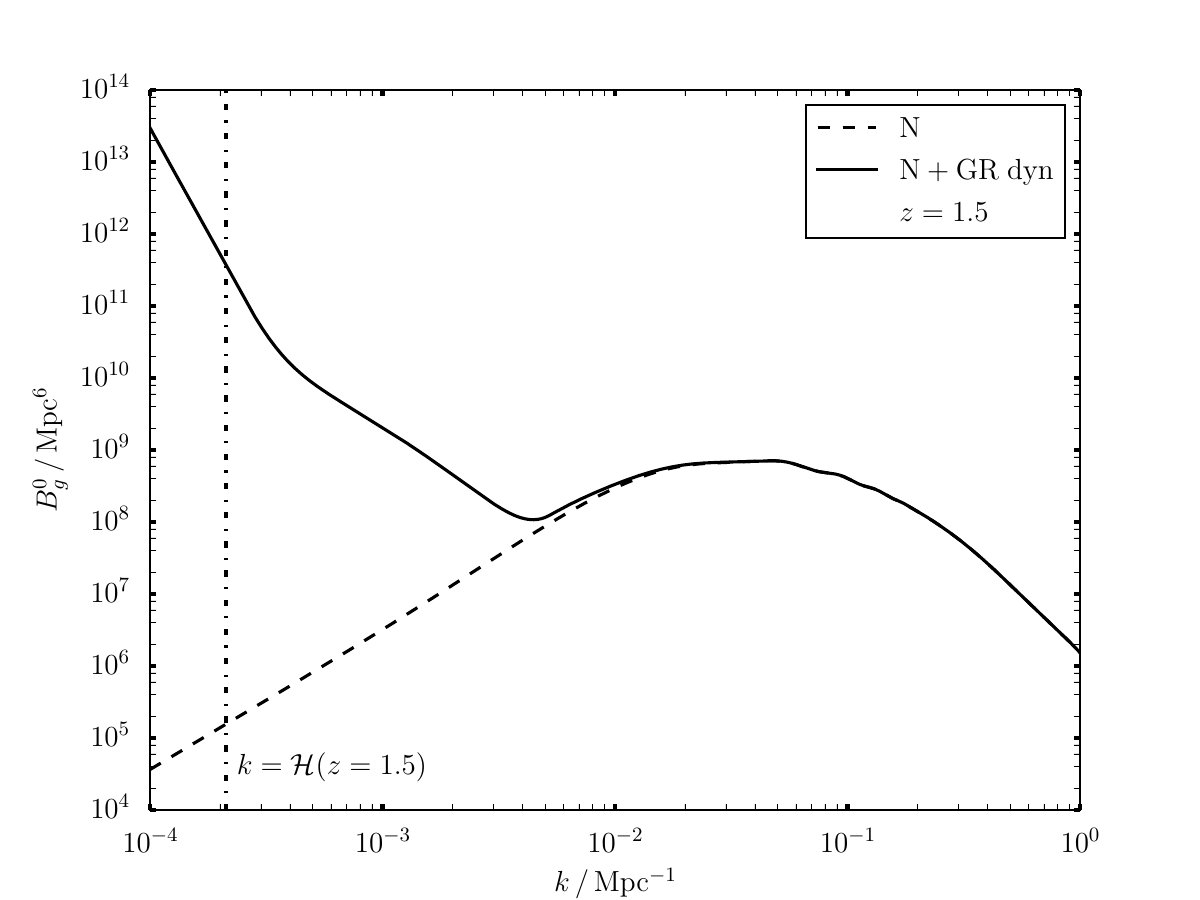}
\caption{\red{The monopole of the galaxy bispectrum in the moderately squeezed configuration \eqref{msc}, at redshifts $z=0.5,1.0,1.5$. The dashed curve is the Newtonian approximation, the solid curve is the result when the {\em only} the relativistic dynamical corrections from \eqref{e15}--\eqref{e19} are included, i.e, only the $\alpha_I$ terms in \eqref{e26}.} 
}
\label{f1}
\end{figure}
\begin{figure}[! ht]
\centering
\includegraphics[width=0.32\textwidth] {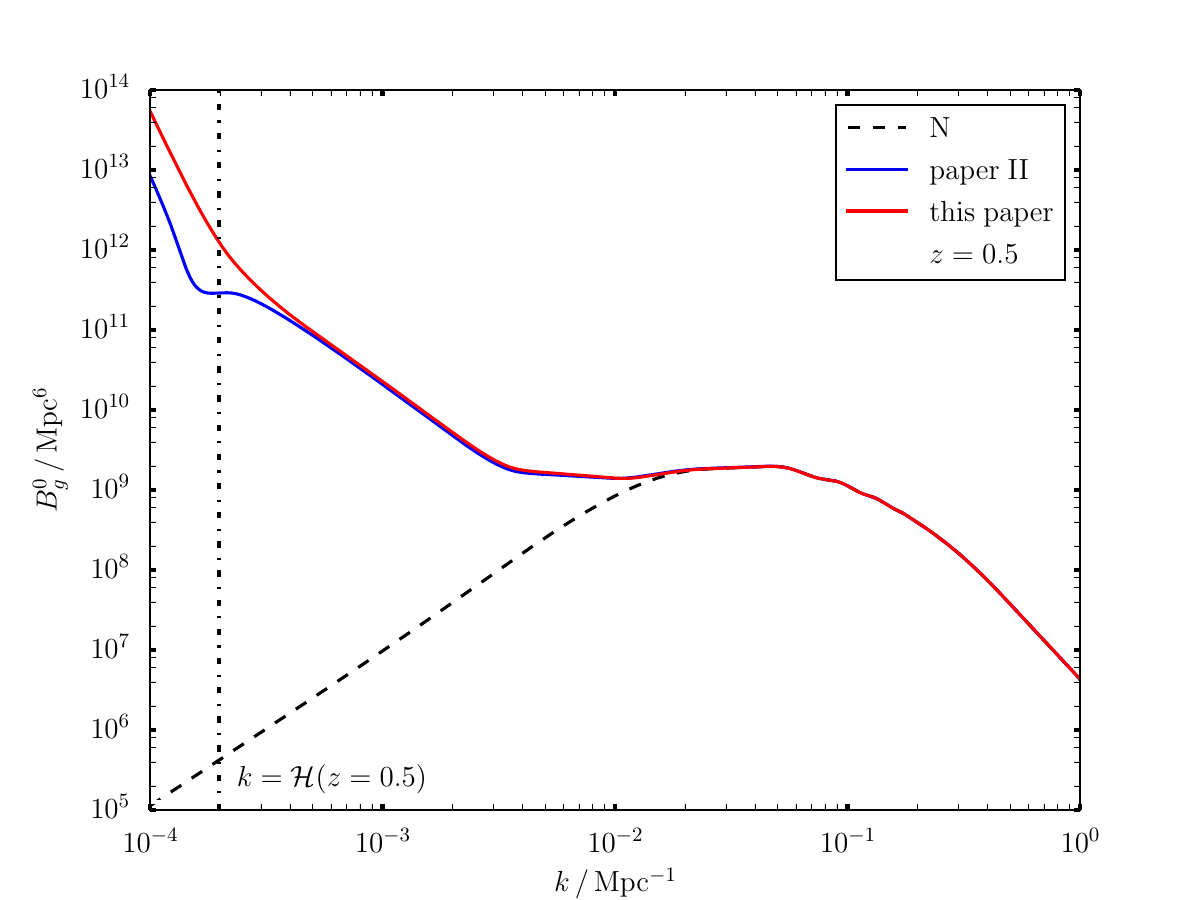}
\includegraphics[width=0.32\textwidth] {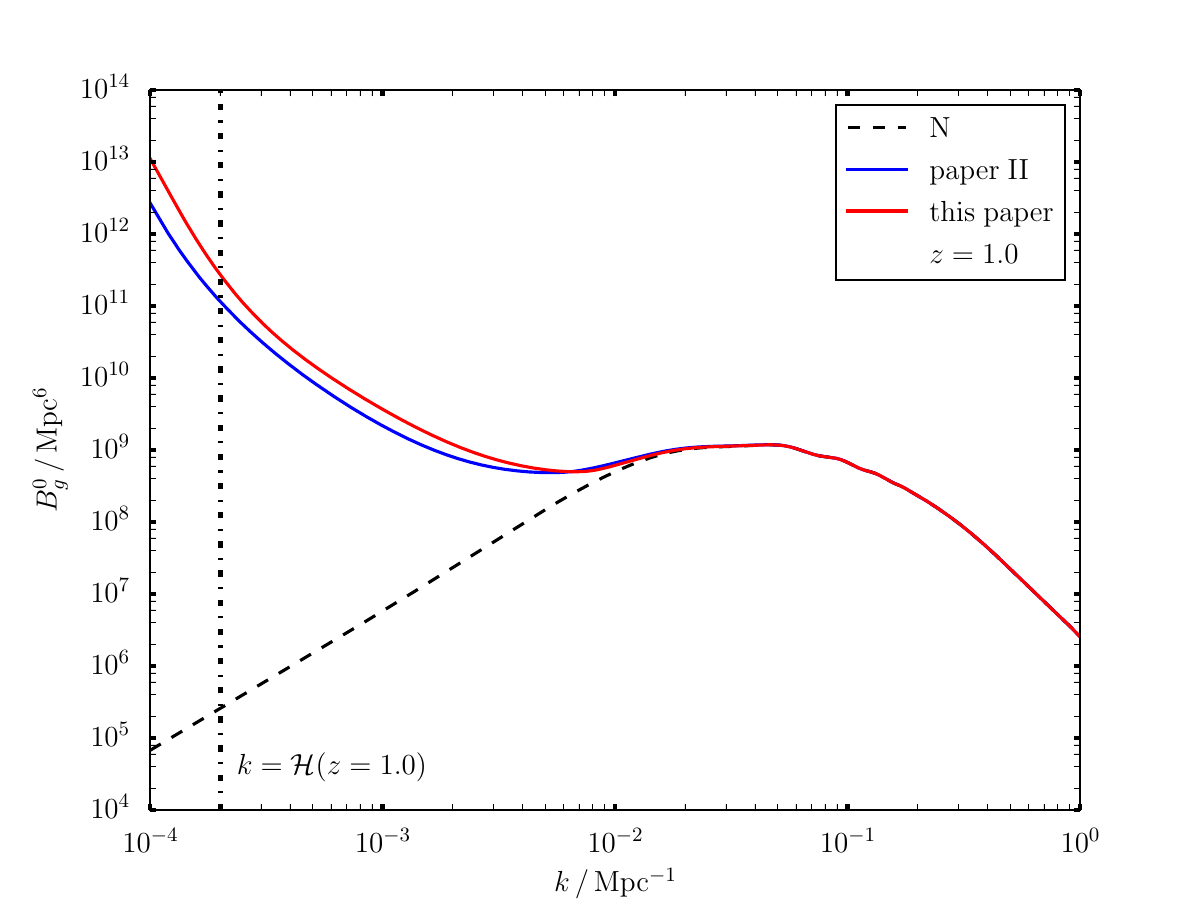}
\includegraphics[width=0.32\textwidth] {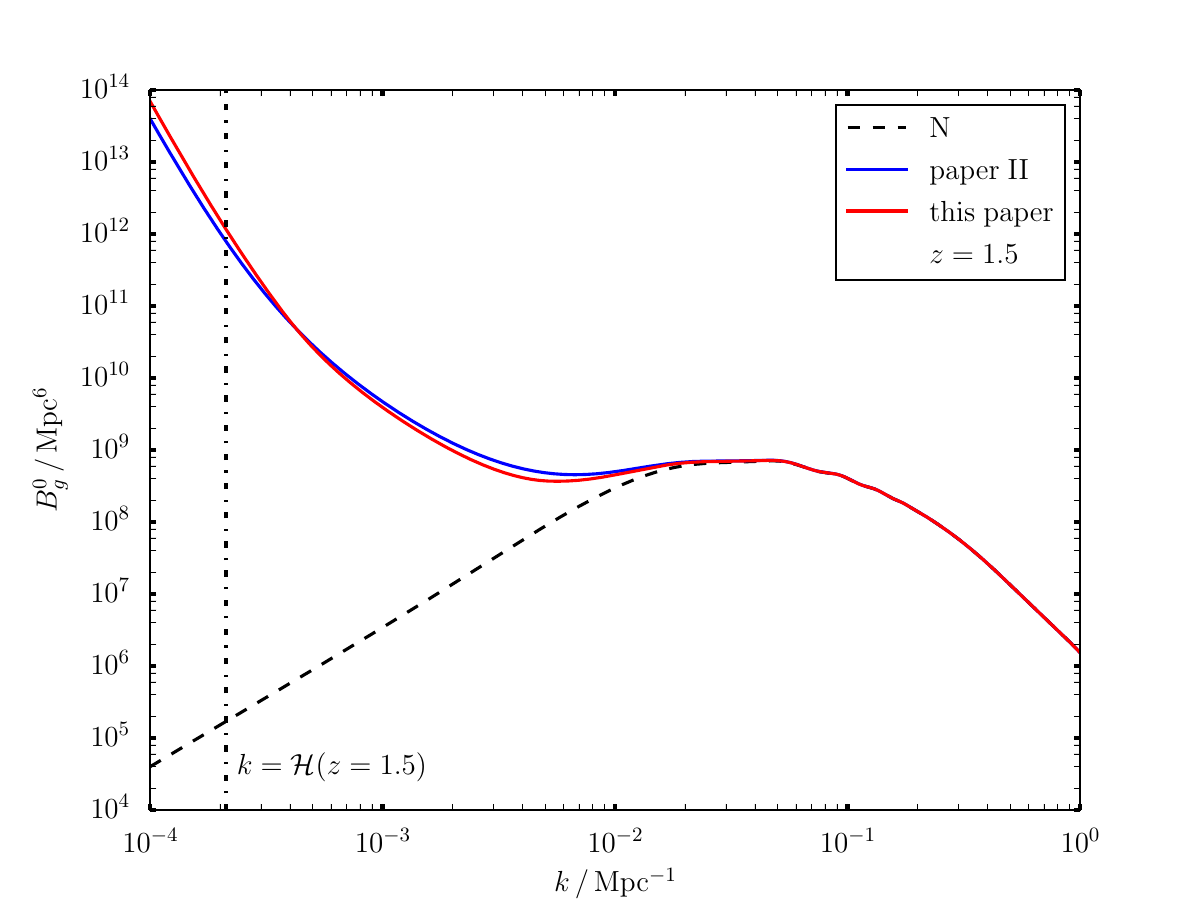}\\
\includegraphics[width=0.4\textwidth] {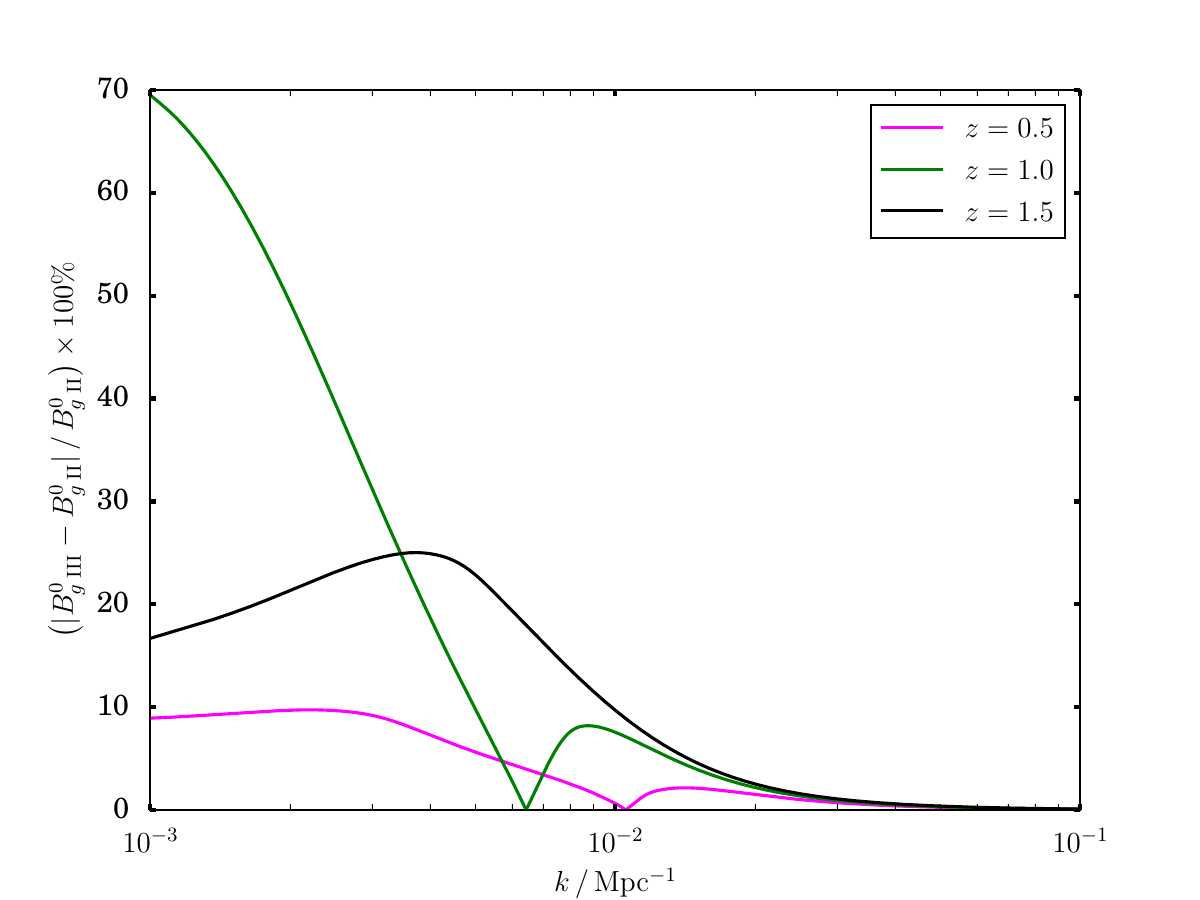}
\caption{\red{{\em Upper:} The monopole of the galaxy bispectrum in the moderately squeezed configuration \eqref{msc}, at redshifts $z=0.5,1.0,1.5$. The dashed curve is the Newtonian approximation, the solid blue curve is the monopole of Paper II~\cite{Jolicoeur:2017nyt} [i.e. only the $\Gamma_I$ terms in \eqref{e26}], and the solid red curve is the full result from \eqref{e26r}, i.e. including relativistic dynamical corrections. \\
{\em Lower:} Percentage difference, at the redshifts $z=0.5,1.0,1.5$, from dynamical corrections to the Paper II monopole.}}
\label{fig:bispectrum_sq1}
\end{figure}
\begin{figure}[! ht]
\centering
\includegraphics[width=0.32\textwidth] {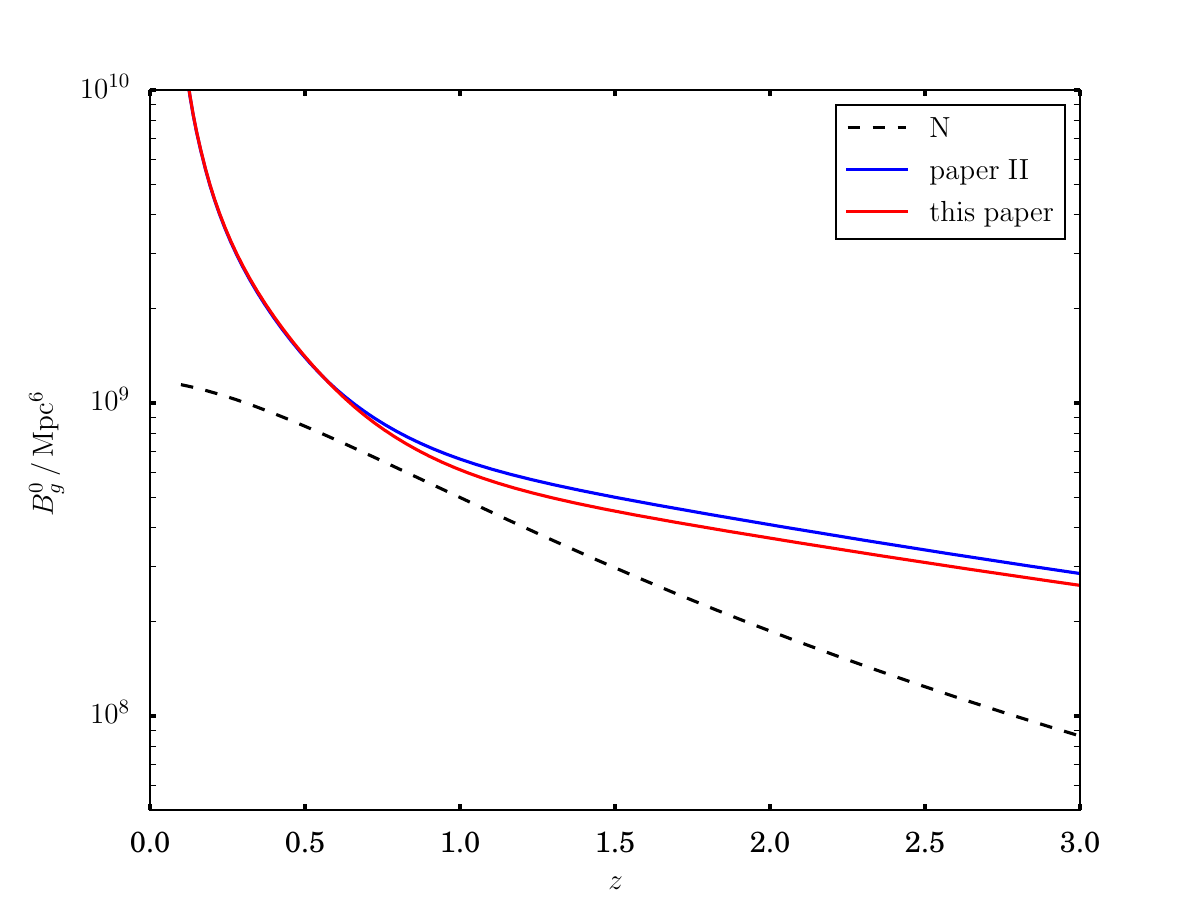}
\includegraphics[width=0.32\textwidth] {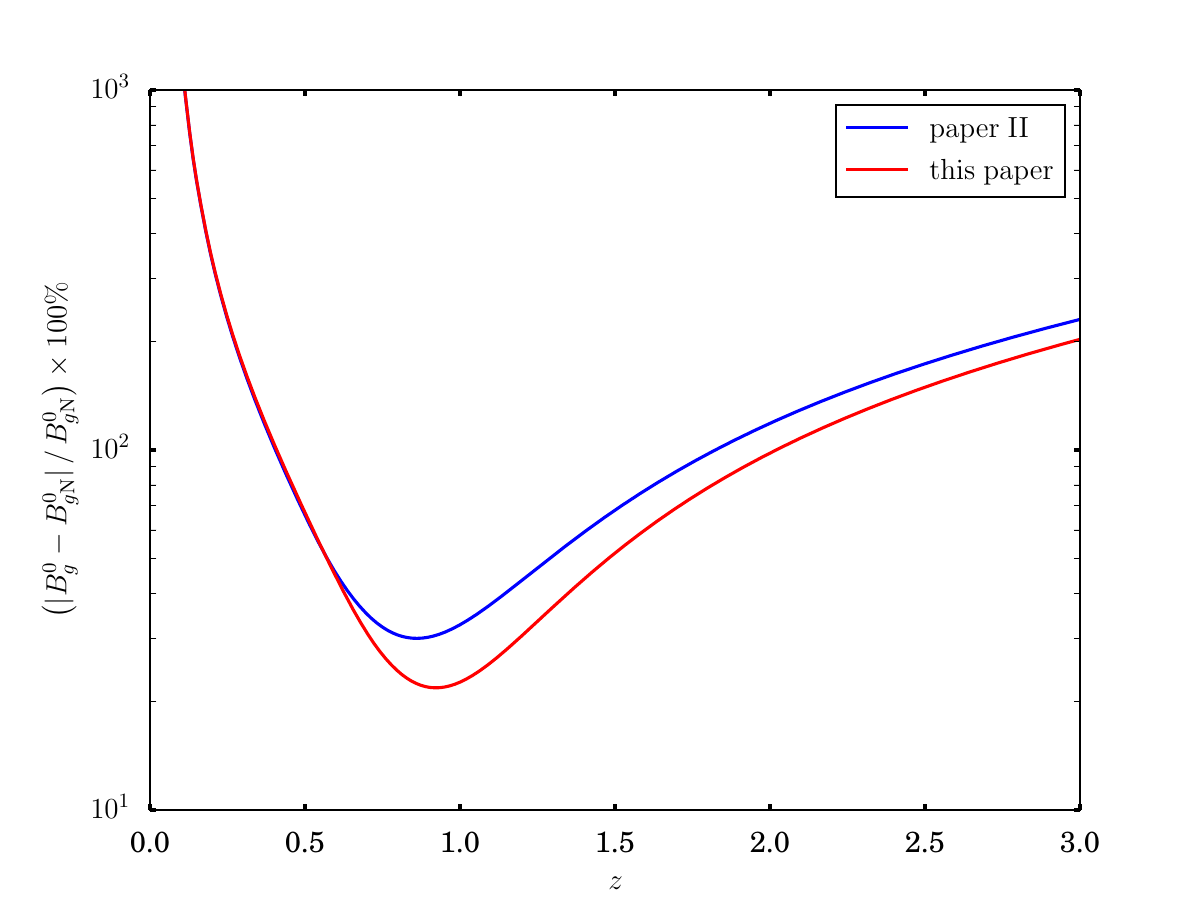}
\includegraphics[width=0.32\textwidth] {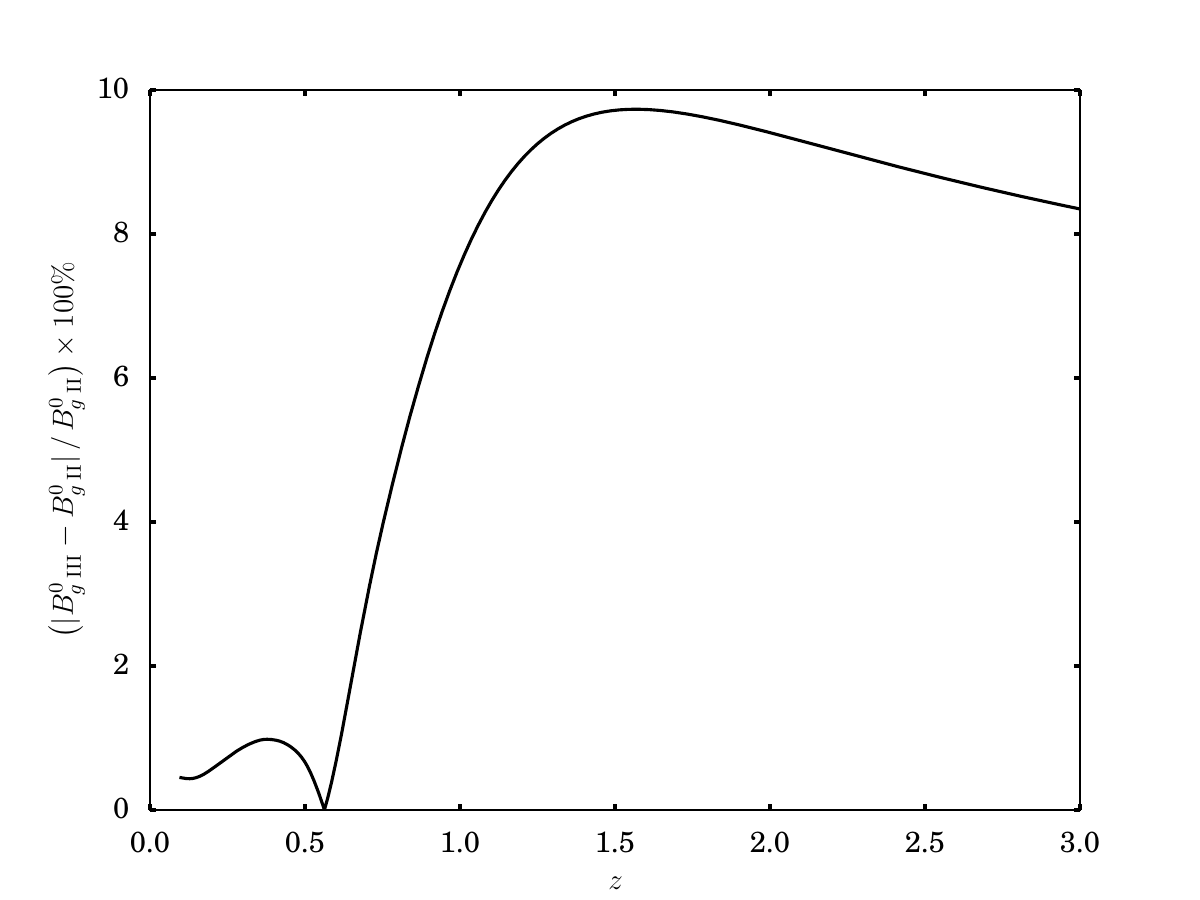}
\caption{ Redshift evolution at  $k = 0.01\;\mathrm{Mpc}^{-1}$ ($\approx$ the equality scale):  the galaxy bispectrum monopole for the Newtonian approximation (dashed),  Paper II (blue) and this paper (red) {\em (left)}; the percentage difference relative to the Newtonian approximation for Paper II and this paper {\em (middle)}; \red{the percentage difference between this paper and Paper II {\em (right)}}.}
\label{f3}
\end{figure}
\begin{figure}[! ht]
\centering
\includegraphics[width=\textwidth] {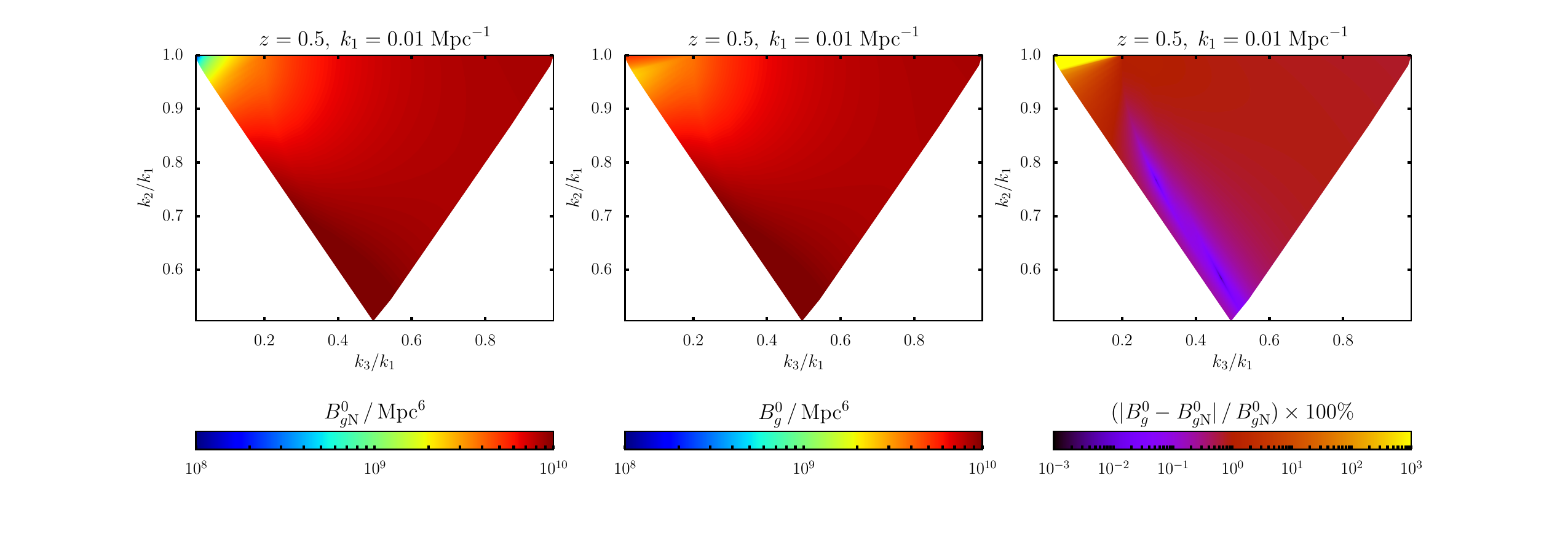} \\
\includegraphics[width=\textwidth]{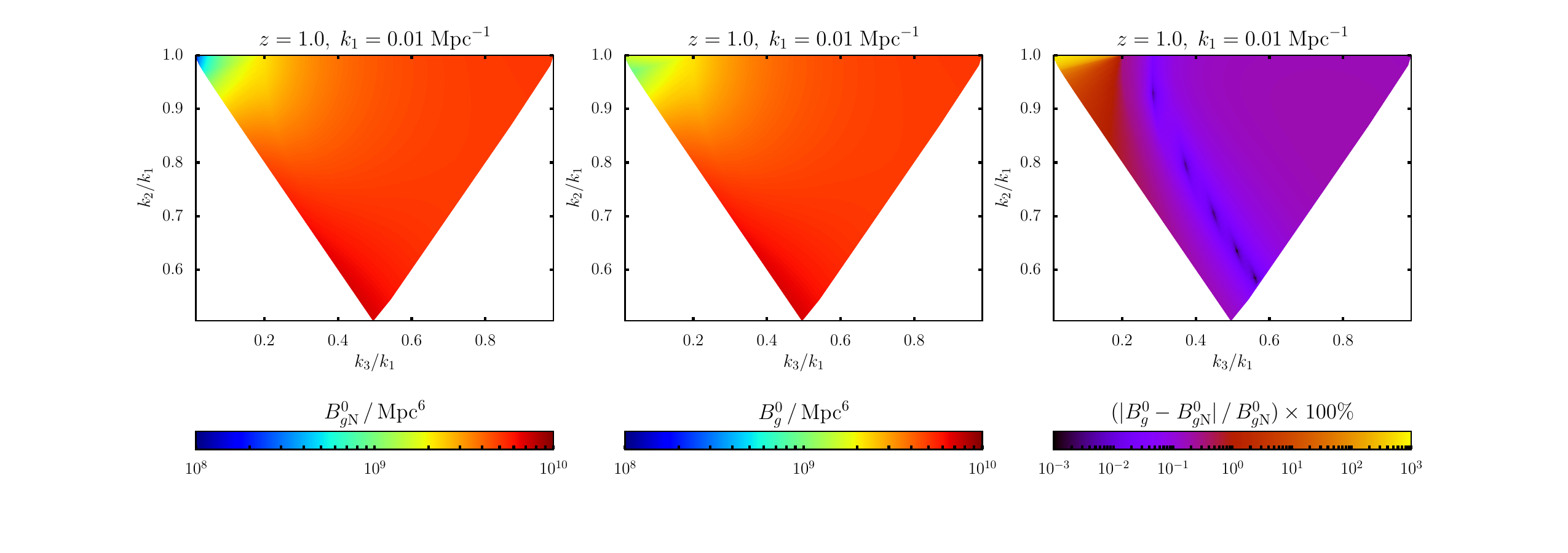} \\
\includegraphics[width=\textwidth]{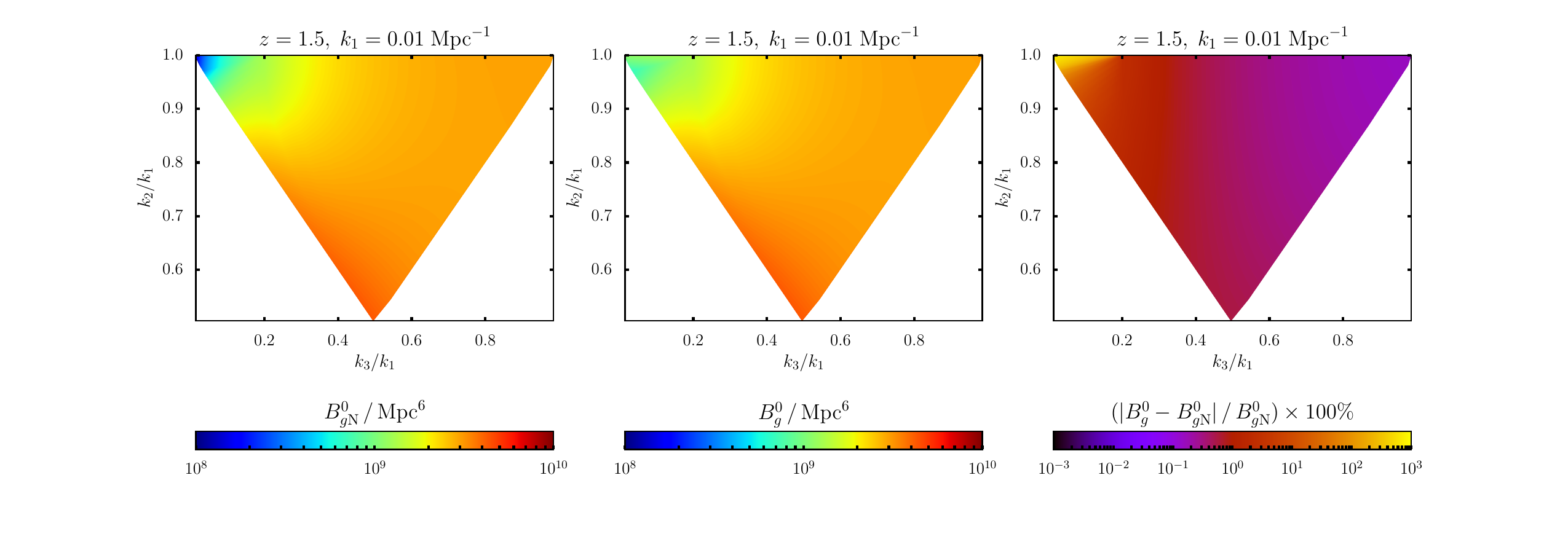}
\caption{ The monopole of the galaxy bispectrum with $k_{1}=0.01\;\mathrm{Mpc}^{-1}$ and at the indicated redshifts: in the Newtonian approximation {\em (left)},  with local projection effects, including dynamical effects, as in this paper {\em (middle)}, and the percentage difference {\em (right)}.  The upper left tip of the wedge is the squeezed limit $(k_{1}=k_{2}, k_{3}\to0)$ and the upper right tip is the equilateral shape $(k_{1}= k_{2} = k_{3})$. }
\label{wedge}
\end{figure}

In order to illustrate quantitatively the impact of the relativistic dynamical corrections \red{on the projection effects presented in Paper II~\cite{Jolicoeur:2017nyt}, we follow Paper II and consider the monopole of the galaxy bispectrum. 
Much of the recent literature, including work on the bispectrum of the BOSS survey~\cite{Gil-Marin:2014sta,Gil-Marin:2016wya}, also uses the monopole of the galaxy bispectrum.} 
\red{As shown in Paper II~\cite{Jolicoeur:2017nyt}, 
$B_g(\k_1,\k_2,\k_3)$ can be written as a function of the magnitudes of the three mode vectors, together with $\mu_{1}$ and $\phi$ (the azimuthal angle around the the line of sight),  
and then  expanded in spherical harmonics: 
\bea
B_g({k}_1,{k}_2,{k}_3,\mu_{1},\phi) &=& \sum_{\ell = 0}
\sum_{m = -\ell}^{\ell}B_g^{\ell m}({k}_1,{k}_2,{k}_3)\, Y_{\ell m }(\mu_1, \phi), 
\\
\label{eq:bispectrummultipoles}
B_{g}^{\ell m}({k}_1,{k}_2,{k}_3) &=& \frac{(2\ell +1)}{4\pi} \int_0^{2\pi} \ud\phi\int_{-1}^{1}\ud\mu_{1}\, B_g({k}_1,{k}_2,{k}_3,\mu_{1}, \phi)\, Y_{\ell m }^{*}(\mu_1, \phi).
\eea
(This can be compared to the Legendre multipole expansion of the galaxy power spectrum.)
The monopole that we consider here is $B^0_g \equiv B^{00}_g$. 

The plane-parallel assumption also affects the monopole, which requires wide-angle corrections. In the Newtonian case, approximate corrections have been applied to the power spectrum (e.g. \cite{Samushia:2015wta,Reimberg:2015jma}) and bispectrum (e.g., \cite{Scoccimarro:2015bla}). In the relativistic case, the full corrections have been explicitly identified in the two-point correlation function by \cite{Bertacca:2012tp,Yoo:2013zga}. 
In principle it is known how to include wide-angle effects in the relativistic galaxy three-point correlation function, but in practice this is very challenging, and has not yet been computed. 
Results from the two-point correlation function indicate that the plane-parallel approximation breaks down
at constant $z$ when the angular separation is $>{\cal O}(10^\circ)$~ \cite{Bertacca:2012tp,Yoo:2013zga}. At
$z\sim 1$ this corresponds to transverse separations  $k<{\cal O}(10^{-3}\,{\rm Mpc}^{-1})$.
This should be kept in mind when considering the plots.}

\red{In order to compare directly with the numerical results of Paper II, we consider the same}  mildly squeezed configuration, with
\be\label{msc}
k_1=k_2\equiv k,~~~ k_3\approx {k\over 16},
\ee
\red{and we choose the same cosmological parameters (from Planck 2015) and the same astrophysical parameters:}
\bea
&& H_0=67.8\,{\rm km/ s/ Mpc},  \qquad \Omega_{m0} =1-\Omega_{\Lambda 0}= 0.308, \\
&& b_1(z)= \sqrt{1+z},~ {b_2(z)=   -0.1 \sqrt{1+z}}, \qquad b_e=0={\cal Q}.
\eea
\red{Note that the vanishing of $b_e$ and ${\cal Q}$ is not physically realistic, but this does not affect our main aim, which is the comparative assessment of the relativistic dynamical corrections to Paper II.}

\red{We start by looking at the effect of the relativistic dynamical corrections on their own, i.e.,
{\em only} the relativistic dynamical corrections from \eqref{e15}--\eqref{e19} are included, which means only the $\alpha_I$ terms in \eqref{e26}, neglecting the relativistic local projection effects of Paper II (from the $\Gamma_I$ terms). Figure~\ref{f1} shows that the dynamical contributions from  the $\alpha_I$ terms, which are of order  $(\cH/k)^n$, $n=2,3,4$, modify the Newtonian approximation significantly from around the equality scale and above, $k\lesssim 0.01\,{\rm Mpc}^{-1}$. The dynamical corrections are qualitatively similar to the projection effects of Paper II. 

In order to quantify the dynamical contribution, we compare the results of
Paper II~\cite{Jolicoeur:2017nyt} [i.e. only the $\Gamma_I$ terms in \eqref{e26}], and full result of this paper, from \eqref{e26r}, i.e. including relativistic dynamical corrections.
The plots of
Fig.~\ref{fig:bispectrum_sq1} (upper panels) compare, for the same redshifts $z=0.5,1,1.5$ as Fig.~\ref{f1}, the monopole from Paper II and from this paper. The lower panel shows the percentage difference in the relativistic monopole from including the dynamical effects, for the scales $10^{-3}<k<10^{-1}\,{\rm Mpc}^{-1}$. It is clearly necessary to include the relativistic dynamical corrections to the projection effects.

Figure~\ref{f3} shows the redshift evolution of the monopole at  $k=0.01\,{\rm Mpc}^{-1}$, i.e. approximately the equality scale, out to redshift $z=3$,  comparing Paper II  with this paper.} 

Finally, we compute the monopole for other configurations, and show the results via a two-dimensional intensity plot in the $(k_2/k_1, k_3/k_1$) plane, where $k_1=0.01\,$Mpc$^{-1}$. To avoid redundancy, we impose $k_1\geq k_2\geq k_3$.
In Fig.~\ref{wedge}, we compare the Newtonian approximation (left panels) with the relativistic bispectrum of this paper (middle panels), at 3 redshifts. The percentage difference between the two is shown in the right panels.

\section{Conclusion}

We extended the results of our previous Paper II~\cite{Jolicoeur:2017nyt}, in which \red{we calculated the local relativistic  lightcone projection effects on the observed galaxy bispectrum, but used the standard Newtonian approximation to compute the contributions of second-order gravitational and velocity potentials.}  There are second-order relativistic corrections from the field equations to these potentials. We computed these dynamical corrections in Fourier space and then derived the ensuing correction to the key second-order kernel of the galaxy bispectrum, which we gave in \eqref{e26r},  together with Appendix~\ref{B}, which gives the coefficients of the updated kernel. 

\red{We used the revised kernel to compute the galaxy bispectrum monopole in some important cases, starting with a mildly squeezed configuration.
Figure~\ref{f1} shows the effect of the relativistic dynamical corrections on their own, i.e.,
with only the relativistic dynamical corrections from the $\alpha_I$ terms in \eqref{e26}, excluding the relativistic local projection effects of Paper II. The dynamical corrections are qualitatively similar to the projection effects of Paper II; in both cases the bispectrum grows on super-equality scales, so that the Gaussian relativistic bispectrum mimics the effect of primordial non-Gaussianity on the Newtonian bispectrum. This suggests that it is indeed important to include the dynamical corrections. 
 
Figure~\ref{fig:bispectrum_sq1} compares the monopole from Paper II [only the $\Gamma_I$ terms in \eqref{e26}] and from this paper, using \eqref{e26r}. The percentage difference in the relativistic monopole from including the dynamical effects is significant on super-equality scales (in the lower panel we used a cut-off $k>10^{-3}\,{\rm Mpc}^{-1}$ to avoid large wide-angle corrections).

We showed in Fig.~\ref{f3} the evolution of the monopole at  $k=0.01\,{\rm Mpc}^{-1}$, up to redshift $z=3$, comparing this paper with  Paper II. At low $z$, the dynamical effects make a sub-percent contribution, but this increases at higher $z$ to $\sim 10\%$. 

In order to consider other configurations, in Fig.~\ref{wedge} we generated a colour intensity map on the wedge $k_1\geq k_2\geq k_3$  in the $(k_2/k_1, k_3/k_1$) plane, with $k_1=0.01\,$Mpc$^{-1}$. This map shows the difference between the full relativistic local bispectrum of \eqref{e26r}, and the Newtonian bispectrum. As expected, the difference is greatest near the squeezed limit.}

This work is part of a series of papers on the observed galaxy bispectrum, starting with the simplest cases and building towards the general case. 
We have not yet included: 
\begin{itemize}
\item
primordial non-Gaussianity;
\item
tidal stress in the galaxy bias;
\item
the second-order effect of the radiation era on 
initial conditions for sub-equality modes;
 \item
integrated contributions to the projection effects, wide-angle correlations and radial (cross-bin) correlations. 
\end{itemize}

\red{The last point involves the highest degree of technical complexity, requiring an angular bispectrum or three-point correlation function analysis on the past lightcone. The first two points are non-trivial:  deriving a relativistic generalisation of the simplest sub-Hubble model of bias (Gaussian and local-in-mass-density, as used in Paper II and this paper) is not easy, and including tidal stress and primordial non-Gaussianity is a significantly more difficult unsolved problem. The third point requires the use of a second-order Einstein-Boltzmann code, as in \cite{Tram:2016cpy}.}

Future papers will address these points.

~\\~\\~\\
\noindent{\bf Acknowledgments:}\\
SJ and RM are supported by the South African SKA Project. OU, RM {and CC} are supported by the UK STFC, {Grants ST/N000668/1 (OU, RM) and  ST/P000592/1 (CC)}.

\newpage
\appendix
\section{\red{Some basic equations and definitions}}\label{AA}

\red{The metric and the 4-velocity of galaxies (equal to the dark matter 4-velocity on the scales of interest) are given in Poisson gauge by
\bea
a^{-2}\,\ud s^2 &= &-\left[1+2\Phi^{{{({1})}}}+\Phi^{{{({2})}}}\right]\ud\eta^2+ \left[1-2\Phi^{{{({1})}}}-\Psi^{{{({2})}}}\right]\ud{\bm x}^2,\\
a\, u^\mu&=&\big(1+v^0,v^i\big),\qquad v^i = \partial^i\Big[ v^{{{({1})}}}+{1\over2}v^{{{({2})}}} \Big],
\eea
where we neglect the vector and tensor modes (see the discussion in Paper II~\cite{Jolicoeur:2017nyt}).

The evolution bias and magnification bias are given by derivatives of the background cumulative luminosity function $\bar{\mathcal{N}}$:
\bea
\label{be}
b_e(a,\bar L) &=&\frac{\partial \ln\big[a^3 \bar{\mathcal{N}}( a , {>}\bar L)\big]}{\partial \ln  a},\\
\label{Q}
\mathcal{Q}(a,\bar L) 
&=&-\frac{ \partial \ln \bar{\mathcal{N}}( a , {>}\bar L)}{\partial\ln \bar L},
\end{eqnarray}
where $\bar L$ is the threshold luminosity of the survey.

The kernel coefficients in \eqref{k1} are \cite{Jeong:2011as}
\begin{eqnarray}
{\gamma_1\over \mathcal{H}} &=&  { f}\bigg[b_{e}  - 2\mathcal{Q} - \frac{2(1-\mathcal{Q})}{\chi\mathcal{H}}- \frac{\mathcal{H}'}{\mathcal{H}^{2}} \bigg] , \label{ga1}
\\
\frac{\gamma_{2}}{\mathcal{H}^{2}} &=& f(3-b_{e}) +\frac{3}{2}\Omega_{m} \left[2+b_{e}-f-4\mathcal{Q}-2 \frac{\left(1- \mathcal{Q} \right)}{\chi\mathcal{H}}-\frac{\mathcal{H}'}{\mathcal{H}^2}\right]. \label{ga2}
\end{eqnarray}

Using the T-gauge real-space expressions given in~\cite{Villa:2015ppa}, we can derive the Fourier-space Eulerian kernels for second-order density contrast and velocity  in \eqref{eq:FourierNewtonian}: 
\begin{eqnarray} 
F_{2}(\bm{k}_{1}, \bm{k}_{2}) &=&1+ \frac{F}{D^2} + \frac{\bm{k}_{1} \cdot \bm{k}_{2}}{k_{1}k_{2}}\bigg(\frac{k_{1}}{k_{2}} + \frac{k_{2}}{k_{1}}\bigg) +\left( 1- \frac{F}{D^2}\right)\bigg(\frac{\bm{k}_{1} \cdot \bm{k}_{2}}{k_{1}k_{2}}\bigg)^{2}  ,
\label{f2} \\
G_{2}(\bm{k}_{1}, \bm{k}_{2}) &=& \frac{F'}{DD'} + \frac{\bm{k}_{1} \cdot \bm{k}_{2}}{k_{1}k_{2}}\bigg(\frac{k_{1}}{k_{2}} + \frac{k_{2}}{k_{1}}\bigg) + \left(2-\frac{F'}{DD'} \right)\bigg(\frac{\bm{k}_{1} \cdot \bm{k}_{2}}{k_{1}k_{2}}\bigg)^{2}, \label{g2}
\end{eqnarray}
where $F$ is a second-order growth factor, given by the growing mode solution of~\cite{Villa:2015ppa}
\be
F''+\cH F'-{1\over \alpha a}F= {1\over \alpha a} D^2.
\ee
These results agree with~\cite{Bernardeau:2001qr}.
In an Einstein-de Sitter background, $F=3D^2/7$ -- and this is a very good approximation in $\Lambda$CDM.}

\newpage
\section{New coefficients in the relativistic second-order kernel \eqref{e26}} \label{A}

The new coefficients in \eqref{e26} are:
\begin{eqnarray}
\frac{\alpha_{1}}{\cH^{4}} &=& \frac{9}{2}\Omega_{m}^{2}\bigg[5 -b_{e} - 2f + \frac{2(1-\Q)}{\chi \cH} + \frac{\cH'}{\cH^{2}}\bigg]+ \frac{3}{2}\Omega_{m}f\bigg[-1 {-f\bigg(4-2f+b_{e}-4\Q-\frac{2(1-\Q)}{\chi\cH}-\frac{4\cH'}{\cH^{2}}\bigg)} \nonumber \\
&&{} + 2b_{e} - 4\Q -\frac{2(1-\Q)}{\chi \cH} - \frac{{3}\cH'}{\cH^{2}}\bigg] - \frac{3}{2}\Omega_{m}\frac{f'}{\cH}\big(1{-2f}\big) - 2f^{2}\big(3+b_{e}\big) \\\nonumber\\
\frac{\alpha_{2}}{\cH^{4}} &=& -\frac{9}{2}\Omega_{m}^{2}\bigg(1-b_{e}+2\Q + \frac{2(1-\Q)}{\chi \cH} + \frac{\cH'}{\cH^{2}}\bigg) - 3\Omega_{m}f\bigg[2 +f\bigg(-1-b_{e} + 2\Q + \frac{2(1-\Q)}{\chi \cH} + \frac{\cH'}{\cH^{2}}\bigg) + b_{e} - 4\Q \nonumber \\
&&{} - \frac{2(1-\Q)}{\chi \cH} - \frac{2\cH'}{\cH^{2}}\bigg] + 3\Omega_{m}\frac{f'}{\cH}
\\ \nonumber \\
\frac{\alpha_{3}}{\cH^{3}} &=& \frac{3}{2}\Omega_{m}f\bigg[-b_{e} + 2\Q + \frac{2(1-\Q)}{\chi \cH} + \frac{2\cH'}{\cH^{2}}\bigg] + 2f^{2}\bigg[-b_{e} + 2\Q + \frac{2(1-\Q)}{\chi \cH}+ \frac{\cH'}{\cH^{2}}\bigg] \\ \nonumber \\
\frac{\alpha_{4}}{\cH^{3}} &=& 3\Omega_{m}f\bigg[b_{e} - 2\Q - \frac{2(1-\Q)}{\chi \cH} - \frac{\cH'}{\cH^{2}}\bigg] \\ \nonumber \\
\frac{\alpha_{5}}{\cH^{2}} &=& -\frac{3}{2}\Omega_{m}f - 2f^{2} \\ \nonumber \\
\frac{\alpha_{6}}{\cH^{2}} &=&3\Omega_{m}f.
\end{eqnarray}

\newpage
\section{Revised coefficients of the full relativistic second-order kernel \eqref{e26r}} \label{B}

By Appendix~\ref{A} and Paper II~\cite{Jolicoeur:2017nyt}, the $\beta_I$ in \eqref{e26r} are:
\begin{eqnarray} 
\frac{\beta_{1}}{\mathcal{H}^{4}} &=& \frac{9}{4}\Omega_{m}^{2}\Bigg[7+4b_e+4{\Q}-4fb_{e}+8f\Q-8b_{e}\Q
+b_{e}^{2}+16\Q^{2}-16\frac{\p \Q}{\p \ln\bar{L}}
 -\frac{2f'}{\cH}+ \frac{b_{e}'}{\cH}-8\frac{\Q'}{\cH} 
 \nonumber\\ 
&&{}+\frac{\cH'}{\cH^{2}}\bigg( 4f-6-2b_e+8\Q+3\frac{\cH'}{\cH^{2}}  \bigg)- \frac{\cH''}{\cH^{3}}
-\frac{2}{\chi\cH}\bigg(2-4f+2b_e-2\Q+4f\Q-2b_e\Q+8\Q^2 
 \nonumber\\ 
&&{}
-8\frac{\p \Q}{\p \ln{\bar{L}}} +3(\Q-1) \frac{\cH'}{\cH^{2}} 
 - 2\frac{\Q'}{\cH}\bigg)
  +\frac{2}{\chi^{2}\cH^{2}}\bigg(1-\Q+2\Q^{2}-2\frac{\p \Q}{\p \ln{\bar{L}}}\bigg)\Bigg]
 \nonumber\\ 
&&{}     
  +{3\over2}\Omega_{m}f\Bigg[13-10f+8b_{e}+fb_{e}-28\Q+2f^2-2b_e^2+4f\Q+8b_e\Q-2\frac{b_{e}'}{\cH}
  +8{\Q'\over\cH}
 \nonumber\\ 
&&{}     
+\frac{2}{\chi\cH}\bigg( f-7+2b_e-f\Q -2b_e\Q-2{\Q'\over\cH} \bigg)\Bigg] 
+f^{2}\Bigg[6-9b_{e}+b_{e}^{2}+\frac{b_{e}'}{\cH}+(b_{e}-3)\frac{\cH'}{\cH^{2}}\Bigg]
 - \frac{3}{2}\Omega_{m}\frac{f'}{\cH} \\
\nonumber\\
\frac{\beta_{2}}{\cH^{4}} &=& -\frac{9}{2}\Omega_{m}^{2}\bigg(1-b_{e}+2\Q + \frac{2(1-\Q)}{\chi \cH} + \frac{\cH'}{\cH^{2}}\bigg) - 3\Omega_{m}f\bigg[2 +f\bigg(-1-b_{e} + 2\Q + \frac{2(1-\Q)}{\chi \cH} + \frac{\cH'}{\cH^{2}}\bigg) + b_{e} - 4\Q \nonumber \\
&&{} - \frac{2(1-\Q)}{\chi \cH} - \frac{2\cH'}{\cH^{2}}\bigg] + 3\Omega_{m}\frac{f'}{\cH}
\\\nonumber\\
\frac{\beta_{3}}{\mathcal{H}^{3}} &=& \frac{9}{4}\Omega_{m}^{2}(f-2+2\Q) + \frac{3}{2}\Omega_{m}f\Bigg[-2 -f\bigg(-3+f+2b_{e}-3\Q-\frac{4(1-\Q)}{\chi\cH}-\frac{2\cH'}{\cH^{2}}\bigg)-\frac{f'}{\cH}+3b_{e}+b_{e}^{2}-6b_{e}\Q+ {4}\Q\nonumber \\
&&{}+8\Q^{2}-8\frac{\p \Q}{\p \ln{\bar{L}}} -6\frac{\Q'}{\cH} +\frac{b_{e}'}{\cH} +\frac{2}{\chi^{2}\cH^{2}}\bigg(1-\Q+2\Q^{2}-2\frac{\p \Q}{\p \ln{\bar{L}}}\bigg) + \frac{2}{\chi \cH}\bigg(-1 -2b_{e}+2b_{e}\Q+\Q-6\Q^{2}  \nonumber \\
&&{} +\frac{3\cH'}{\cH^{2}}(1-\Q) +6\frac{\p \Q}{\p \ln{\bar{L}}} + 2\frac{\Q'}{\cH}\bigg) -\frac{\cH'}{\cH^{2}}\bigg(3+2b_{e}-6\Q-\frac{3\cH'}{\cH^{2}}\bigg) - \frac{\cH''}{\cH^{3}}\Bigg] + f^{2}\Bigg[-3+2b_{e}\bigg(2+\frac{(1-\Q)}{\chi\cH}\bigg)\nonumber \\
&&{}-b_{e}^{2}+2b_{e}\Q -6\Q-\frac{b_{e}'}{\cH}-\frac{6(1-\Q)}{\chi\cH}+2\bigg(1-\frac{1}{\chi\cH}\bigg)\frac{\Q'}{\cH}\Bigg] \\\nonumber\\
\frac{\beta_{4}}{\cH^{3}} &=& \frac{3}{2}\Omega_{m}f\bigg[-b_{e} + 2\Q + \frac{2(1-\Q)}{\chi \cH} + \frac{2\cH'}{\cH^{2}}\bigg] + 2f^{2}\bigg[-b_{e} + 2\Q + \frac{2(1-\Q)}{\chi \cH}+ \frac{\cH'}{\cH^{2}}\bigg], \\ \nonumber\\
\frac{\beta_{5}}{\cH^{3}} &=& 3\Omega_{m}f\bigg[b_{e} - 2\Q - \frac{2(1-\Q)}{\chi \cH} - \frac{\cH'}{\cH^{2}}\bigg] \\\nonumber\\
\frac{\beta_{6}}{\cH^{2}} &=& \frac{3}{2}\Omega_{m}\Bigg[2-2f+b_{e}-4\Q-\frac{2(1-\Q)}{\chi\cH}-\frac{\cH'}{\cH^{2}}\Bigg] \\\nonumber\\
\frac{\beta_{7}}{\cH^{2}} &=& f(3-b_{e}) \\\nonumber\\
\frac{\beta_{8}}{\cH^{2}} &=& 3\Omega_{m}f(2-f-2\Q) + f^{2}\Bigg[4+b_{e}-b_{e}^{2}+4b_{e}\Q-{6}\Q-4\Q^{2}+4\frac{\p \Q}{\p \ln{\bar{L}}} + 4\frac{\Q'}{\cH} - \frac{b_{e}'}{\cH}  \nonumber \\\nonumber\\
&&{} - \frac{2}{\chi^{2}\cH^{2}}\bigg(1-\Q+2\Q^{2}-2\frac{\p \Q}{\p \ln{\bar{L}}}\bigg) - \frac{2}{\chi\cH}\bigg(3-2b_{e}+2b_{e}\Q-\Q-4\Q^{2}+\frac{3\cH'}{\cH^{2}}(1-\Q) + 4\frac{\p \Q}{\p \ln{\bar{L}}} + 2\frac{\Q'}{\cH}\bigg) \nonumber \\\nonumber\\
&&{} - \frac{\cH'}{\cH^{2}}\bigg(3-2b_{e}+{4}\Q+\frac{3\cH'}{\cH^{2}}\bigg) + \frac{\cH''}{\cH^{3}}\Bigg]  
\\\nonumber
\end{eqnarray}
\begin{eqnarray}
\frac{\beta_{9}}{\cH^{2}} &=& -\frac{3}{2}\Omega_{m}f - 2f^{2} \\ \nonumber \\
\frac{\beta_{10}}{\cH^{2}} &=& 3\Omega_{m}f
\\\nonumber\\
\frac{\beta_{11}}{\cH^{2}} &=& 3\Omega_{m}f - f^{2}\Bigg[-1+b_{e}-2\Q- \frac{2(1+\Q)}{\chi\cH}-\frac{\cH'}{\cH^{2}}\Bigg] \\\nonumber\\
\frac{\beta_{12}}{\cH^{2}} &=& \frac{3}{2}\Omega_{m}\Bigg[b_{1}\bigg(2+b_{e}-4\Q-\frac{2(1-\Q)}{\chi\cH} -\frac{\cH'}{\cH^{2}}\bigg) + \frac{b_{1}'}{\cH} + 2\bigg(2-\frac{1}{\chi\cH}\bigg)\frac{\p b_{1}}{\p \ln{\bar{L}}}\Bigg] - f\Bigg[b_{1}(f-3+b_{e}) + \frac{b_{1}'}{\cH}\Bigg]  \\\nonumber\\
\frac{\beta_{13}}{\cH^{2}} &=& \frac{9}{4}\Omega_{m}^{2} + \frac{3}{2}\Omega_{m}f\Bigg[{1}-2f+2b_{e}-{6}\Q-\frac{4(1-\Q)}{\chi\cH}-\frac{3\cH'}{\cH^{2}}\Bigg] + f^{2}(3-b_{e}) \\ \nonumber\\
\frac{\beta_{14}}{\cH} &=& -\frac{3}{2}\Omega_{m}b_{1} \\\nonumber\\
\frac{\beta_{15}}{\cH} &=& 2f^{2}\\\nonumber\\
\frac{\beta_{16}}{\cH} &=& f\Bigg[b_{1}\bigg(f+b_{e}-2\Q-\frac{2(1-\Q)}{\chi\cH}-\frac{\cH'}{\cH^{2}}\bigg) + \frac{b_{1}'}{\cH} + 2\bigg(1-\frac{1}{\chi\cH}\bigg)\frac{\p b_{1}}{\p \ln \bar{L}}\Bigg] \\\nonumber\\
\frac{\beta_{17}}{\cH} &=& -\frac{3}{2}\Omega_{m}f \\\nonumber\\
\frac{\beta_{18}}{\cH} &=& \frac{3}{2}\Omega_{m}f -f^{2}\Bigg[3-2b_{e}+{4}\Q+\frac{4(1-\Q)}{\chi\cH}+\frac{3\cH'}{\cH^{2}}\Bigg] \\\nonumber\\
\frac{\beta_{19}}{\cH} &=& f\Bigg[b_{e}-2Q-\frac{2(1-\Q)}{\chi\cH}-\frac{\cH'}{\cH^{2}}\Bigg]
\end{eqnarray}

\newpage
\bibliographystyle{JHEP}
\bibliography{reference_library}

\end{document}